\documentclass{aa}
\usepackage{amssymb}
\usepackage{amsmath}
\usepackage{graphicx}
\usepackage{color,xcolor}
\usepackage[varg]{txfonts}
\usepackage{natbib}
\usepackage[colorlinks=true,citecolor=blue,urlcolor=blue,linkcolor=blue]{hyperref}
\usepackage[utf8]{inputenc}

\DeclareMathOperator{\sinc}{sinc}

\begin{document}

\title{The coherent magnetic field of the Milky Way halo, Local Bubble and Fan Region}
\author{Alexander Korochkin$^{1}$, Dmitri Semikoz$^{2}$, and Peter Tinyakov$^{1}$}
\institute{Université Libre de Bruxelles, CP225 Boulevard du Triomphe, 1050 Brussels, Belgium \email{alexander.korochkin@ulb.be}
\and Université de Paris Cite, CNRS, Astroparticule et Cosmologie, F-75013 Paris, France
}
\authorrunning{A. Korochkin et al}
\titlerunning{GMF}

\abstract
{Recent catalog of Faraday rotation measures (RM) of extragalactic sources, together with the synchrotron polarization data from WMAP and Planck, provides us with a wealth of information on magnetic fields of the Galaxy. However, the integral character of these observables together with our position inside the Galaxy makes the inference of the coherent Galactic magnetic field (GMF) complicated and ambiguous.}
{We combine several phenomenological components of the GMF --- the spiral arms, the toroidal halo, the X-shaped field and the field of the Local Bubble --- to construct a new model of the regular GMF outside the thin disk.}
{We use the binned $\chi^2$ approach to fit the parameters of the model to the data. To have control over the relative contributions of the RM and polarization data to the fit we pay special attention to the estimation of errors in data bins. To this end we develop a systematic method which is uniformly applicable to different data sets. This method takes into account individual measurement errors, the variance in the bin as well as fluctuations in the data at angular scales larger than the bin size. This leads to decrease of the errors and, as a result, to better sensitivity of the data to the model content. We cross checked the stability of our method with the new LOFAR data which have tiny errors of the measurements of individual sources.}
{We found that the four components listed above are sufficient to fit both the RM and polarization data over the whole sky with only a small fraction masked out. Moreover, we have achieved several important improvements compared to previous approaches. Due to account of our location inside of the Local Bubble our model does not require introduction of striated fields. For the first time we showed that the Fan Region can be modeled as a Galactic-scale feature. The pitch angle of the magnetic field in our fit converged to the value around 20 degrees. Interestingly, this value is very close to the direction of the spiral arms inferred recently from Gaia data on upper main sequence stars.}
{}


\maketitle

\section{Introduction}
It is well known that our Galaxy is permeated with magnetic field. Such a field is intricately linked to the evolution of the galaxy, affecting gas dynamics and, in particular, influencing the star formation, propagation of cosmic rays, and shaping the Galactic emission at radio frequencies. Accurate account of the Galactic magnetic field (GMF) is one of the keys for solving the long standing puzzle of sources of ultra-high-energy cosmic rays (UHECR). However, the structure, origin and evolution of the GMF is still not well understood.

Apart from the short-scale turbulent field, radio observations of external galaxies indicate the presence of a large-scale coherent field ordered on scales comparable to the size of the galaxy \citep{2015A&ARv..24....4B}. Interestingly, two essentially different coherent components can be identified. For the first component, the magnetic field lies in the plane of the galactic disk and forms so called magnetic arms, which are not necessarily coincident with the gaseous or optical spiral arms. These spiral magnetic structures are found in almost every spiral galaxy observed so far (e.g., magnetic arms of M51 \citep{2011MNRAS.412.2396F} or the disk field of M101 \citep{2016A&A...588A.114B}). The second component, on the contrary, extends out of the disk into the galactic halo and forms an X-shaped pattern which can be traced up to the distances of several kpc away from the galactic plane (e.g., the halo field of NGC~4631 \citep{2013A&A...560A..42M} or NGC~5775 \citep{2011A&A...531A.127S}).

In the standard paradigm, the galactic coherent field is explained as a result of action of large-scale dynamo (for a review see \citet{Beck:2019jyi} and \citet{2023ARA&A..61..561B}) which can also link together the coherent field of the disk and the X-shaped field of the halo \citep{2008A&A...487..197M}. Alternatively, it can be the result of the gravitational compression of a primordial magnetic field of nanogauss strength occurred during structure formation \citep{Howard:1996qx}.

While other galaxies are observed from outside and their global magnetic field structure can be relatively easily recognized, our location inside the Galactic disk makes such measurements for the Milky Way a non-trivial problem. Typically used tracers of magnetic field such as Faraday rotation measures (RM) of polarized sources or synchrotron emission of cosmic ray electrons (CRE) give information integrated over the line of sight. Thus, one has to solve the inverse problem to reconstruct the global 3-dimensional GMF structure from sky projections of several GMF-sensitive observables. This is not possible without assumptions about the global GMF structure. 

The situation is further complicated by the fact that these integral tracers get contributions from magnetized and ionized bubbles located in the Galactic disk \citep{2010ApJ...724L..48W} typically within $100 - 200$~pc from the Galactic plane. In fact, the solar system itself is located {\em inside} one of such bubbles usually referred to as the Local Bubble \citep{LocalBubble_2}, which is generally believed to be remnant of supernova explosions. Several other nearby bubbles seen from the exterior can also be identified. They produce large-scale anomalies in the RM and synchrotron polarization data at high Galactic latitudes that have to be subtracted or masked. The problem gets worse as more bubbles contribute for the lines of sight closer to the Galactic plane, and appears untractable without invoking distance-sensitive (tomographical) methods. As a result, the information on the global structure of the magnetic field in the disk, particularly in the direction of central Galaxy, is very uncertain.

Several phenomenological models of the coherent GMF have been proposed previously which differ by complexity, underlying data used, and fitting techniques. First generation of models used synchrotron data and earlier compilations of 
rotation measures of pulsars and extragalactic sources \citep{1997A&A...322...98H, Tinyakov:2001ir, 2001SSRv...99..243B}. After the arrival of the first all-sky rotation measure catalog of extragalactic objects \citep{NVSS} the next generation of models included it as the main data source for reconstruction of the magnetic field in the halo \citep{Pshirkov:2011um}. The combination of the rotation measures with the synchrotron polarisation data gives an additional powerful tool for model building \citep{JF_GMF_1}.  Recent models using various combinations of the existing data include \citet{Han2018, Han2019, Shaw:2022lqd, Unger:2023lob,Han2024}. 

The purpose of this paper is to present a new model of the GMF with a focus on the halo field, by which we understand in what follows the field outside of the thin disk. There are several motivations to update the existing models. On one hand, a significant amount of new RM data has recently appeared. For instance, the S-PASS/ATCA survey \citet{2019MNRAS.485.1293S} filled the blind spot of the RM catalog \citet{NVSS} used in previous studies. The Wilkinson Microwave Anisotropy Probe (WMAP) synchrotron polarization data have been updated in \citet{WMAP_23GHz} and more recently in \citet{Watts:2023vdc}. On the other hand, even the most recent existing models leave a lot of room for improvement. The model of \citet{Pshirkov:2011um} did not include the synchrotron polarization data. The model of \citet{JF_GMF_1} assumed the magnetic field to be to some degree "striated" (flipping direction in random domains)  which is essentially a way to boost the synchrotron polarization signal with respect to  the rotation measures by an arbitrary factor treated as a phenomenological parameter. The last but not the least, there is a large freedom in choosing and parameterizing the main components of the GMF. Comparing the results of independently constructed fits gives some understanding of the systematic uncertainties involved. While this paper was in preparation there appeared a new model of the GMF by \citet{Unger:2023lob}; their results and those presented here should be considered complementary in this sense.

In this paper we use the WMAP synchrotron polarization  measurements at 23~GHz \citep{WMAP_23GHz} and the catalog of about 59000 RMs of extragalactic sources \citep{2023ApJS..267...28V} to fit the parameters of our model. Even though mostly the same data were used in previous models including the recent model by \citet{Unger:2023lob}, we have made several essential improvements. 

First, we model and take into account the contribution of the Local Bubble. Its importance  was first pointed out by \citet{LocalBubble_1} when considering polarized dust emission observed by Planck at 353~GHz. In previous models that used the RM and synchrotron data together it was found that the magnetic field preferred by the RM data alone is insufficient by a factor of 1.5-3 \citep{Unger:2023lob} to explain the observed polarized intensity at high Galactic latitudes. To reconcile the two sets of data the "striation" parameter was introduced. Here we show that a sizeable contribution of the Local Bubble to the polarization signal at 23~GHz solves this "synchrotron deficit" problem while being fully compatible with the RM data. Using the simplest model of the Local Bubble, we achieved a significant improvement in the quality of the fit, especially at high Galactic latitudes $|b|>60^\circ$, thus completely eliminating the need for the "striated" magnetic fields. 

Second, we have developed a uniform approach to the error estimation which, although technical, is a crucial step in fitting together data of a different nature, the RM and synchrotron polarization in our case. We therefore have control of a relative importance of the RM and synchrotron polarization data in the global fit. The key observation here is that one of the main origins of the "noise" in the data is common in both cases --- the magnetized bubbles in the disk. In general, our error bars are smaller than obtained by previously used methods. 

Third, we found from our fit that the pitch angle of the local magnetic arms is  $20^\circ$. This pitch angle is significantly larger than found in previous studies where the pitch angle did not exceed $\sim 10^\circ$. This large value is in perfect agreement with the pitch of the Galactic arms inferred from Gaia data by \citet{gaia}, as is illustrated below in Fig. \ref{fig:spiral_arms}. 

Finally, combining magnetic field in the Local arm with that of the Perseus arm we, for the first time, obtained a good fit of the Fan region --- a large bright region on the synchrotron polarization maps roughly in the direction of the Galactic anti-center. According to recent studies \citep{fan} the polarization signal in this region is produced at $1-2$ kpc from the solar system and thus makes part of the global structure (for the model of the Fan Region as a local feature see \citet{2021ApJ...923...58W}). In previous GMF models this region was masked as a local anomaly. 

The rest to of the paper is organized as follows. In the Sect.~\ref{sec:data} we describe our method for calculation of errorbars of rotation measure and synchrotron data and discuss data we used. In Sect.~\ref{sec:model} we describe our model for Galactic Magnetic Field in the halo. In Sect.~\ref{sec:global_fit} we present the results of the global fit of our model to the data. In Sect.~\ref{sec:discussion} we discuss the performance of our model. In particular, in Sect.~\ref{sec:LocalBubble} we go in details of new feature introduced in the present model for the first time, namely the Local Bubble. Finally we sum up our results in the Sect.~\ref{sec:conclusions}. The Appendices~\ref{appendix:cleaning} and~\ref{appendix:Derivation} are devoted to the method of error assignment. In the Appendix~\ref{appendix:comparison} we compare our results to the previous models. 

Throughout the paper and in the numerical code, we fix the coordinate system such that the center of the Galaxy is at the origin, the $z$-axis is directed towards the Galactic North pole, and the solar system coordinates are \{-8.2~kpc, 0, 0\}. For the observer at Earth the direction towards positive $y$ is $(l,b) = (90^\circ, 0^\circ)$, towards positive $z$ -- $(l,b) = (0^\circ, 90^\circ)$.  

\section{Data preparation}
\label{sec:data}
Galactic magnetic field can be probed with different tracers. Among them are polarized thermal emission of dust, Faraday rotation measures of Galactic and extragalactic sources and synchrotron emission of cosmic ray electrons at different frequencies. Each of these tracers is sensitive to different properties of magnetic field but all of them provide us with valuable information about coherent GMF structure. Below we will briefly review each tracer and justify the choice of datasets for our analysis.

\paragraph{Dust.} Observations of dust polarized emission reveal the average orientation of the component $B_\perp$ of GMF perpendicular to the line of sight, without providing any information on its strength. The dust is concentrated in the thin disk with the height $\sim 100$~pc, thus making the method mainly sensitive to our local neighborhood. Significant progress has been achieved thanks to the Planck measurements at a frequency of 353 GHz, where dust emission dominates. Firstly, it was shown that none of the existing models can accurately predict dust emission \citep{2016A&A...596A.103P}. Secondly, it was demonstrated that the emission at high Galactic latitudes is dominated by the Local Bubble wall \citep{2019A&A...631L..11S, LocalBubble_2}. Moreover, it was found that a model where the field in the wall is obtained by compressing an initially uniform field successfully explains the structure of polarization after fitting the direction of the initial field. We adopted this approach when building our model of the Local Bubble, see Sect.~\ref{sec:model}.

\paragraph{Rotation measures: Galactic pulsars.} Faraday rotation is the phenomenon of rotation of the polarization plane of a linearly polarized radiowave when propagating through the magnetized plasma. The rotation angle $\Delta \theta$ is proportional to the square of the wavelength $\lambda$; the coefficient is called the rotation measure (RM): 
\begin{equation} \label{eq:rot}
    \Delta \theta = {e^3\lambda^2\over 2\pi m_e^2}  \int_0^l n_e B_{||} ds \equiv
    \mathrm{RM}\cdot\lambda^2 ,
\end{equation}
where $e$ and $m_e$ are the electron charge and mass, and the integral is taken along the line of sight between the observer and the source. The RM is proportional to the density of free electrons $n_e$ and the component of the magnetic field parallel to the line of sight $B_{||}$,
\begin{equation} \label{eq:RM}
    \mathrm{RM} \approx 0.812\,\,\int_0^l \left[\frac{n_e(s)}{\mathrm{cm^{-3}}}\right] \left[\frac{B_{||}(s)}{10^{-6}\,\mathrm{G}}\right] \,\left[\frac{\mathrm{d}s}{\mathrm{pc}}\right]\,\,\,\,\mathrm{rad/m^2}.
\end{equation}  
Positive RM corresponds to the field pointing towards the observer. Measurement of the polarization on at least two different wavelengths allows one to calculate RM. 
\begin{figure}
    \centering
    \includegraphics[width=0.99\linewidth]{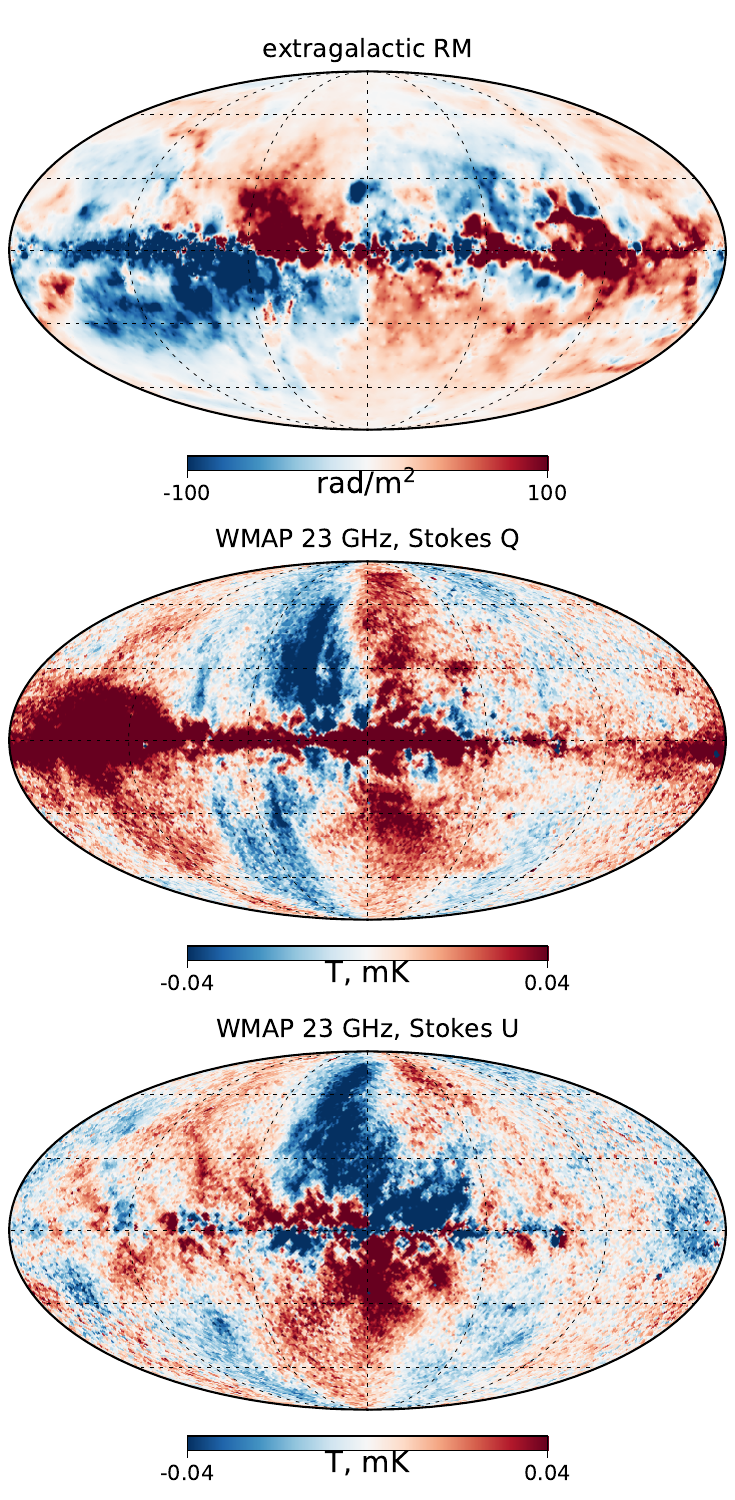}
    \caption{Compilation of the data used in this paper. Upper panel: Extragalactic RM skymap, smoothed and cleaned by~\citet{rm2020}. Note that in our main analysis, we start with the individual RMs from the CIRADA catalog. The RM skymap from \citet{rm2020} is shown for clearer visual representation. Middle and bottom panels: WMAP synchrotron Stokes Q and U skymaps at~23~GHz \citep{WMAP_23GHz}.}
    \label{fig:data}
\end{figure}

Pulsars are bright sources of polarized radio emission. The advantage of pulsars is their negligibly small intrinsic RM values. Also their location {\em within} the galaxy allows for the investigation of the inner structure of the GMF. Study of pulsar RMs gives an indication of the magnetic field reversal in the direction to the Galactic center \citep{reversal}. However, pulsars are mostly concentrated in the disk, which makes them not so useful for studying the GMF halo. Moreover, the distances to pulsars are usually inferred from their dispersion measures, which introduces additional uncertainties. For these reasons we do not use pulsar data in this paper and leave the validation of our model against pulsars for the future.

\paragraph{Extragalactic RM.} At the moment more than 59000 extragalactic RMs have been measured which are summarized in publicly available master catalog by \citet{2023ApJS..267...28V}. In our study we use the latest compilation of this catalog, namely CIRADA consolidated catalog version v1.2.0 \footnote{The catalog can be found online at the link \url{https://github.com/CIRADA-Tools/RMTable}}. The core of the catalog is the NRAO VLA Sky Survey (NVSS) survey \citep{NVSS}. Additionally, it comprises RMs from \citet{Schnitzeler2019, VanEck2021, Betti2019, Farnes2014, Mao2010, Tabara1980, Simard-Normandin1981, Broten1988, Riseley2020, Brown2003, Taylor2024, Mao2012LMC, Mao2012Halo, Feain2009, VanEck11, Ma2020, OSullivan2017, Kaczmarek2017, Anderson2015, Brown07, Klein2003, Heald09, Shanahan2019, Clarke2001, Minter1996, Ranchod2024, VanEck2018a, Law2011, Riseley2018, Livingston2022, Mao2008, Roy2005, Livingston2021, Oren1995, Clegg1992, Kim2016, Battye2011, Ma2019, Rossetti2008, Costa2018, Vernstrom2018, Gaensler2001, Costa2016}, including the most recent extensive Low-Frequency Arrar (LOFAR) survey LoTSS \citep{OSullivan2023} and the Australian Square Kilometre Array Pathfinder survey POSSUM \citep{Vanderwoude2024}.

The totality of extragalactic RMs covers the whole sky almost uniformly with roughly 1 source per 1 deg$^2$ on average, which is very convenient for studying the global structure of the GMF. The smoothed and cleaned RM skymap constructed by \citet{rm2020}\footnote{Note that the RM skymap of \citet{rm2020} while cleaned of outliers is not suitable for our purposes since the correlations between individual pixels of the map do not allow for the estimation of the errorbars of larger bins, see Sec~\ref{subsec:bins_and_errors}.} from the previous version of the catalog (v0.1.8) is shown in Fig.~\ref{fig:data}, upper panel. An even earlier version of the catalog was used for the first time to study magnetic field in the Galactic halo by \citet{Pshirkov:2011um}.

Unlike pulsars, extragalactic sources generally do have intrinsic RMs which sometimes may be very large, so the RM data have to be cleaned from these outliers. The exact procedure is described in the Appendix~\ref{appendix:cleaning}. The bulk of the extragalactic sources have intrinsic RMs with the zero mean and the r.m.s. not exceeding $\sim 7$~rad/m$^2$ \citep{Schnitzeler:2010ax,Pshirkov:2013wka}. 

\paragraph{Synchrotron.} Relativistic electrons of energy $E$ spiraling in the magnetic field emit radio waves near the critical frequency
\begin{equation} \label{eq:crit_freq}
    \nu_c \approx 1.6\,\left[\frac{B_\perp}{10^{-6}\,\mathrm{G}}\right]\left[\frac{E}{10\,\mathrm{GeV}}\right]^2\,\,\,\mathrm{GHz}.
\end{equation}
The resulting synchrotron radiation is polarized in the plane perpendicular to the direction of the magnetic field. Assuming the distribution of cosmic ray electrons (CRE) in energy is known, the information about $B_\perp$ can be extracted from synchrotron measurements. This information is complementary to that contained in RMs. 

The CMB experiments WMAP and Planck performed the full-sky synchrotron polarization measurements with high precision at 23~GHz and 30~GHz, respectively. The WMAP synchrotron skymap was first combined with RM data to study the GMF in \citet{2009JCAP...07..021J, 2010RAA....10.1287S, JF_GMF_1,JF_GMF_2}. 

At higher frequencies the polarized emission is dominated by the emission from dust. On the other hand, at lower frequencies the Faraday rotation becomes important as can be seen from equations~\ref{eq:rot} and \ref{eq:crit_freq}. Significant Faraday rotation leads to depolarization of initially polarized beam. The combination of Faraday rotation with the synchrotron measurements at large number of nearby frequencies allows for the so-called Faraday tomography \citep{1966MNRAS.133...67B}, when the synchrotron emissivity is determined as a function of RM accumulated along the line of sight. Recently such measurements were performed by GMIMS collaboration in the Northern sky in the frequency band 1280 - 1750~MHz \citep{2021AJ....162...35W}, and in the Southern sky in the range 300 - 480~MHz \citep{2019AJ....158...44W}.

In the current study we use the final nine-year WMAP 23~GHz dataset to fit the model. We choose it because firstly, it covers the entire sky, and secondly, it is measured at sufficiently high frequency so that depolarization is negligible. The use of frequencies at which depolarization is significant implies an increasingly significant role of local structures which we do not model in this study. On the other hand WMAP 23~GHz and Planck 30~GHz data are very similar after rescaling for the difference in frequency. We could use either one; we choose WMAP because its seven-year version has already been used in \citet{JF_GMF_1}. The residual systematic difference after rescaling between WMAP and Planck was studied in detail in the Cosmoglobe project \citep{Watts:2023vdc}. We note that this difference is small compared to the current precision of the GMF models, and thus is not important for our purposes. Figure~\ref{fig:data} shows the sky maps of Stokes Q and U parameters (middle and bottom panels, respectively) corresponding to the WMAP 23~GHz dataset. 

\subsection{Binning and error estimation} 
\label{subsec:bins_and_errors}
In order to extract maximum benefit from the combination of the RM and synchrotron polarization data, we prepare both datasets in a similar way. This includes four steps: removal of outliers, masking, binning and error estimation. The procedures for outliers and masking are different for RMs and synchrotron skymaps and are discussed separately in the subsections~\ref{subsec:RM} and~\ref{subsec:synch}. On the other hand, for both datasets we adopt the same binning scheme and errors estimation algorithm which are described below.

After getting rid of outliers and applying the masks we are left with a collection of small pixels (or individual RMs) that can be directly used for modelling. However, as we are interested in the coherent part of the GMF it is useful to average the masked RM and synchrotron data over the angular bins of a relatively large size. Indeed, the signal of coherent GMF is expected to vary on scales of tens of degrees, and so bins of a smaller, but not much smaller angular size are sufficient to trace it. Moreover, averaging over large bins effectively cancels small-scale fluctuations in the data and reduces errorbars. 
\begin{figure}
    \centering
    \includegraphics[width=0.99\linewidth]{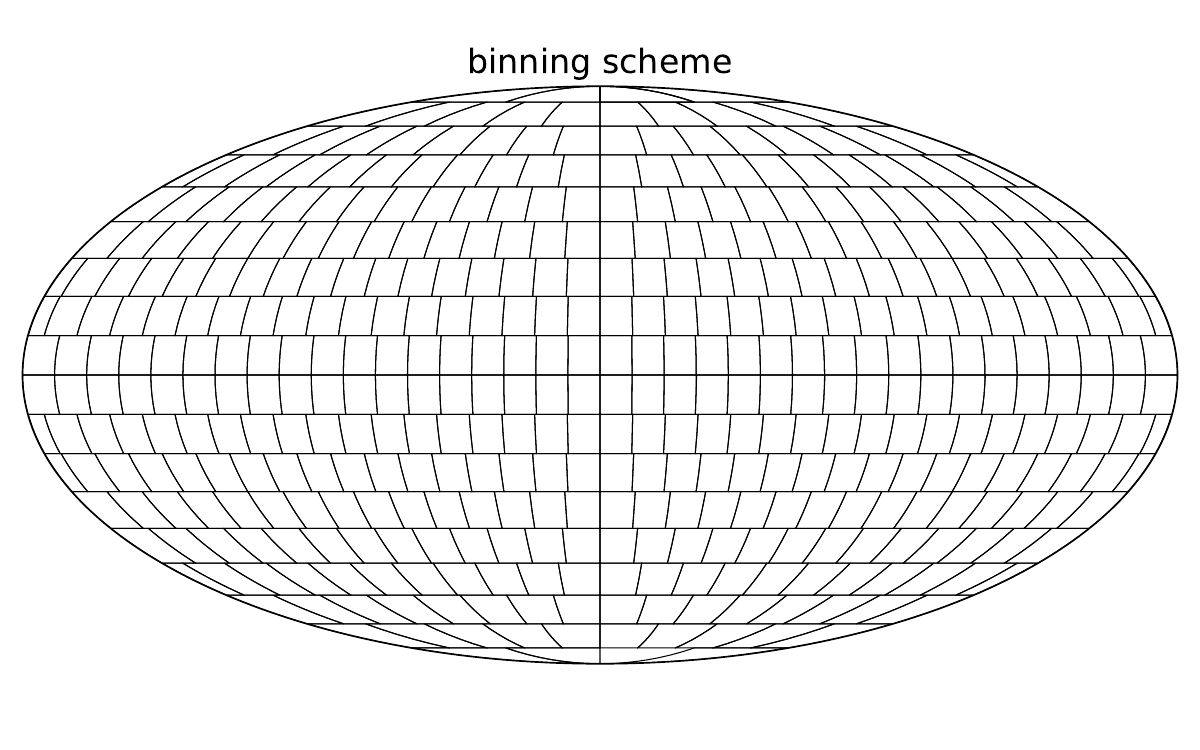}
    \caption{The binning scheme used in this analysis.
     The bins are rectangular in Galactic coordinates and are arranged in iso-latitude bands of $10^\circ$ width. The two equatorial bands are divided into 36 square $10^\circ\times 10^\circ$ bins each. The number of bins in other bands is chosen in such a way as to get the closest approximation to equal-area bins.}
    \label{fig:binning}
\end{figure}

The same approach was already used in previous studies. \citet{Pshirkov:2011um} and \citet{Ferriere2} used the bins with the size $\sim 10^\circ$, \citet{JF_GMF_1, Unger:2023lob}, used smaller bins. Since there is no strict criterion for choosing the optimal bin size we use the same bins as \citet{Pshirkov:2011um} for both RM and synchrotron data, see Fig.~\ref{fig:binning}. The advantage of this binning scheme is that bins form isolatitude bands. The binning scheme was chosen independently of the masks. Those bins where more than 50\% of the original content (by area for WMAP and by the number of sources for RM) has been masked were discarded. 

After setting the binning scheme the most important step is the estimation of the mean value in each bin and assigning corresponding error bars. \citet{JF_GMF_1} and \citet{Unger:2023lob} used the standard deviation within the bins as an error estimate. We do not see  a justification for this choice. If the variations of individual sources (or pixels in the case of polarization data) from the mean were all independent, the error in the bin should be the standard deviation divided by the square root of the number of sources, typically much smaller than the standard deviation itself. However, bubbles and other structures of large angular size (e.g. Radcliffe Wave \citep{2024arXiv240603765P}), which are also part of the noise, give coherent contributions to individual bins, so the errors should be larger, but likely not as large as the standard deviation. \citet{Pshirkov:2011um} used a more complex error estimation procedure which for many bins resulted in smaller errors equal to the standard deviation divided by three. Such a procedure, however, lacks a statistical justification and thus is difficult to generalize to the case of two data sets of a different nature. 
\begin{figure}
    \centering
    \includegraphics[width=0.99\linewidth]{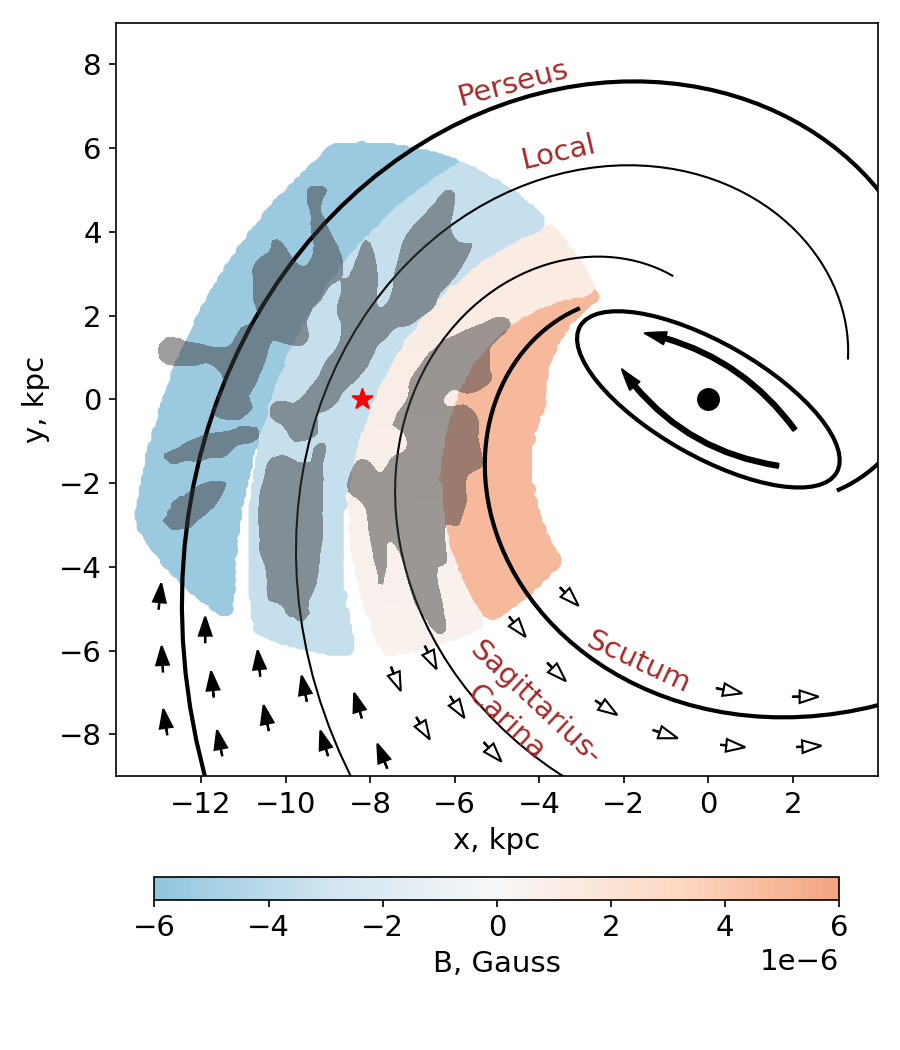}
    \caption{Schematic picture of the Galactic arms as viewed from the North Galactic pole. Black spiral curves shows the axes of the best-fit spiral magnetic arms of our model. The intensity of the color indicates the magnetic field strength in the Galactic plane $z=0$. The blue (orange) colored regions correspond to the magnetic field directed inward/clockwise (outward/counter-clockwise) as marked by the black arrows at the bottom of the plot. Black dot marks the Galactic center while an ellipse represents the Galactic bar. Red star marks the position of the Sun. Grey regions show the spiral arms as deduced from  Gaia observations. The pitch angle in the solar vicinity is $20^\circ$. }
    \label{fig:spiral_arms}
\end{figure}

In this work we propose a new method of error estimation. The observed quantities --- RMs or polarized intensity maps --- are sums of contributions from the coherent GMF and from fluctuations of different origin, including inhomogeneities of the interstellar medium (ISM), the turbulent component of the GMF, magnetized or ionized clouds, intrinsic RMs of sources, etc., and of different scales including those exceeding the bin size. We do not model such fluctuations and consider them as noise obscuring the coherent GMF signal. Our method to estimate the contribution of this noise to the errors assigned to bins is based on the assumption that its statistical properties only depend on the latitude $b$. We assess the parameters of the noise at given $b$ by the Fourier analysis of the data in the isolatitude strips, which we then use to estimate the errors.

If the data in a strip only contained noise, then the variance associated with a bin of size $L$ would be given by the sum over modes of the power spectrum with the window function $\sinc^2(kL/2)$, see  Appendix~\ref{appendix:Derivation} for the derivation. Note that this variance includes contributions from scales larger than the bin size. The actual data also contain the coherent part which is concentrated in the lowest modes and should be subtracted. In fact, the model that we will fit to the data only has sizeable power in the first three harmonics $k=0,1,2$, so we subtract these harmonics before calculating the variance, see Fig.~\ref{fig:strip_fourier}. Specifically, for all bins of the same strip we calculate the variance $\sigma_\mathrm{L}^2$ according to the formula: 
\begin{equation}
    \sigma^2_\mathrm{L} = 2\sum_{k_0}^\infty \, \sinc^2\left(\frac{kL}{2}\right) S_k
    \label{eq:sigma2_L}
\end{equation}
where $S_k$ is the power spectrum of the strip of data to which the bin belongs, and the bin size $L$ is of order 10$^\circ$. Eq.~\ref{eq:sigma2_L} results in errors that are smaller than the simple data variance over the bin by a factor 2-3. Note also that this approach allows us to treat the RM and synchrotron polarization data on equal footing. 
\begin{figure}
    \centering
    \includegraphics[width=0.99\linewidth]{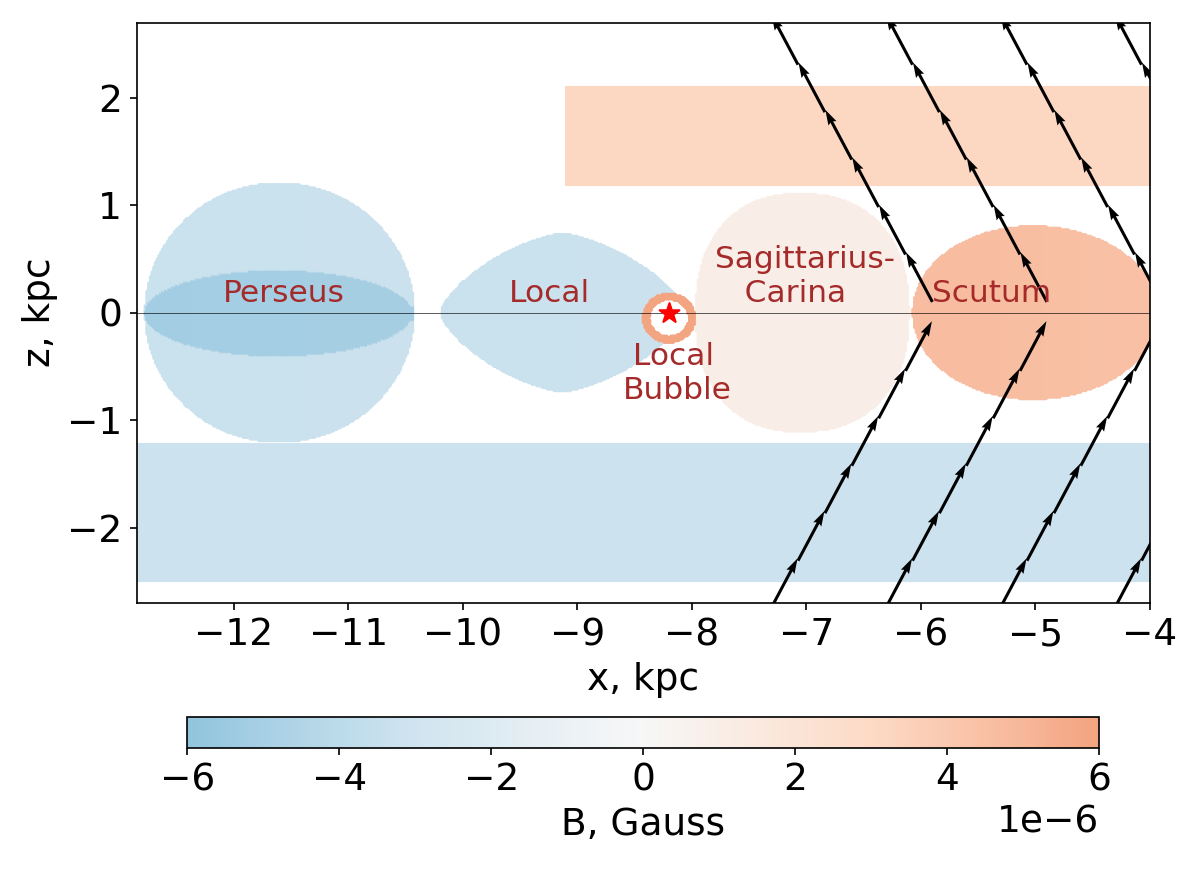}
    \caption{Section of the best-fit GMF model by the plane $y=0$, which is perpendicular to Galactic plane and come through Sun (red star) and Galactic center. Blue and orange regions are the magnetic arms, northern and southern toroidal field and the Local Bubble. The color map is the same as in Fig.~\ref{fig:spiral_arms}. Black arrows show the direction of the X-shaped field. Red star marks the position of the Sun.}
    \label{fig:GMF_xz}
\end{figure}

Eq.~\ref{eq:sigma2_L} assumes that the mean value in a given bin can be calculated exactly, while in reality this is not the case: the data only sample the underlying function in a number of points, so the mean calculated from these values has an associated statistical error $\sigma_\mu$. We determine this error by fitting the distribution of values in the bin by a Gaussian, which gives both the mean and its statistical error: $\mu \pm \sigma_\mu$. The total error in the bin is thus
\begin{equation}
    \sigma_\text{tot} = \sqrt{ \sigma^2_\mathrm{L} + \sigma_\mu^2 }.  
\end{equation}
For the polarization data, and for most (but not all) of the RM bins the contribution of $\sigma_\mu$ is negligible. 

We note that the errors $\sigma_\mathrm{L}$ and $\sigma_\mu$ have different behavior when increasing the number of observations $N$. The error on the mean value of the bin $\sigma_\mu$ tends to zero approximately as $\sim$~1/$\sqrt{N}$, while $\sigma_\mathrm{L}$ remains constant. 

The details of the calculation of $\sigma_\mu$ and $\sigma_\mathrm{L}$ are described in the Appendices~\ref{appendix:cleaning} and~\ref{appendix:Derivation}, respectively. The procedure is the same for both RM and synchrotron datasets.

\subsection{RM mask} \label{subsec:RM}
The value of the RM in some directions may be dominated by  local features of the ISM. Thus, when fitting the coherent GMF they should be masked out. An illustrative example is the well-known Sh2-27 cloud in the direction $(l,b) \approx (10^\circ, 25^\circ)$, ionized by the massive star Zeta Ophiuchi passing through it. This cloud is located at the distance of $\sim$ 100~pc and is clearly visible as a $\sim 10^\circ$ bright blue spot in the RM map, see Fig.~\ref{fig:data}. The average RM in the direction of Sh2-27 goes below $-100$ while RMs in its vicinity are close to zero \citep{2019MNRAS.487.4751T}. 

Since RM is proportional to the density of thermal electrons, potentially biased regions are expected to correlate with highly ionized nearby bubbles. In order to mask them out it is convenient to use maps of H$\alpha$ emission. Similar approach was recently used by \citet{Unger:2023lob}. Using the full-sky H$\alpha$ map compiled by \citet{Finkbeiner:2003yt} we masked pixels where the intensity of H$\alpha$ emission is at least twice as strong as the average intensity at the same latitude.

Additionally, based on visual inspection of the RM map from \cite{rm2020} (see Fig.~\ref{fig:data}) we masked out the arc-like segment with positive RMs around $(l, b) \approx (100^\circ, 0^\circ-50^\circ)$ which is probably related to the Loop III. Also we removed from the analysis bright bubble-like structure at $(l, b) \approx (160^\circ, -15^\circ)$ and a region with mostly zero RM surrounded by strongly negative RMs near $(l, b) \approx (110^\circ, -10^\circ)$.

Finally, we excluded the Galactic plane ($|b| < 10^\circ$) because its modelling depends on the enhancement of thermal electrons density in the spiral arms. Unfortunately, existing models of thermal electrons rely on pre-Gaia measurements of the spiral arms whose pitch was believed to be significantly smaller than that inferred from the Gaia data \citep{gaia}. Our preliminary fits, as well of the final result, favor the same pitch as from Gaia, thus making these models incompatible with our setup. The description of the adopted model of thermal electrons is given in Sect.~\ref{sec:therm_elec}. 

The resulting mask is shown in Fig.~\ref{fig:skymaps}. In total the RM dataset mask covers 26\% of the sky. 

\subsection{Synchrotron mask} \label{subsec:synch}
The synchrotron mask is built independently of the RM mask. First, we discarded all pixels of the WMAP Stokes Q and U maps marked as corrupted. Then we masked out the brightest, most probably nearby features. Namely, we excluded Loop I and Loop II by cutting out corresponding ring segments. The mask is purely geometrical and manually set to cover the majority of Loop I and Loop II. In contrast with previous studies we do not mask the Galactic plane and the Fan region. We apply this mask for both Stokes Q and U skymaps, see Fig.~\ref{fig:skymaps}. The mask covers 11\% of the sky.

There are many other less prominent loops and spurs visible in the polarization data apart from the Loop I and Loop II, see~\citet{2015MNRAS.452..656V}. As these features are not so bright, we do not mask them out in order to keep the mask as small as possible.

\section{Model of the Galactic Magnetic Field}
\label{sec:model}
Sky maps of both Faraday rotation and synchrotron polarization data reveal large regions of constant sign that are aligned with the Galactic plane and the direction to the Galactic center, leaving no doubt of the existence of the coherent magnetic field in the Galaxy. These regions form a particular pattern which has to be reproduced by the magnetic field model. Even though we only measure the projections onto the sky plane of 3-dimensional structures, this requirement turns out to be rather restrictive if one assumes that the model consists of a few simple components. We found that 4 major components are sufficient to describe the global structure of the Faraday rotation and polarization maps: the thick disk, the toroidal halo, the X-shape halo field and the field of the local bubble. Different parts of the data are more sensitive to different components, but all four are needed for a good global fit. In this section we overview these components and their role in reproducing the key features of the data, and present the results of the global fit.

\subsection{Magnetic field components}
\label{sect:components}
\paragraph{Thick disk.} While most of Galactic stars are concentrated within a couple hundred parsecs from the Galactic plane where they populate the Galactic arms, thermal electrons and coherent magnetic fields are usually assumed to form a larger structure often referred to as the {\em thick disk} which generally traces the spiral arms and extends up to 1-2 kpc from the Galactic plane. We assume in this work that the magnetic field associated with this structure is directed along the arms either towards or away from the Galactic center. The solar system is situated at the inner edge of the Local arm; the "local" magnetic field is directed clockwise as viewed from the Galactic North pole, with a pitch towards the center. The detailed structure of the spiral arms in the solar vicinity has been mapped out by Gaia \citep{gaia} and will be discussed further in Sect.~\ref{sec:global_fit}. There are indications from the pulsar data of the field reversal at $\sim 0.5$~kpc in the direction to the Galactic center \citep{reversal}. We therefore assume that the field changes direction to counter-clockwise in the inner Sagittarius-Carina arm. 

The typical thickness of the disk is $\sim 1$~kpc, which means that only the region within $\sim 6$~kpc from the Sun contributes to the observables at Galactic latitudes $|b|>10^\circ$. Correspondingly, only the arms that pass in the relative vicinity of the Sun are constrained from our fits to the data. These are, in the first place, the Local and Perseus arms in the outer Galaxy and the Sagittarius-Carina and Scutum arms in the inner Galaxy. Our fits are not sensitive to the field in the Norma and Outer arms. The general shape of the thick disk field is shown in Fig.~\ref{fig:spiral_arms} where the colored region corresponds to distances within $6$~kpc from the Sun. 

In our model each arm is an independent GMF component. The position of an arm is given by its axis, represented by the logarithmic spiral:
\begin{equation}
    x = -x_0 + a e^{k(\phi + b)} \cos{\phi},
    \quad y = -y_0 + a e^{k(\phi + b)} \sin{\phi},
\end{equation}
where $k=\tan \alpha$ is the tangent of the pitch angle $\alpha$,  $b$ is an initial phase, and $x_0$ and $y_0$ are the origin of the spiral. The cross section of an arm in the plane perpendicular to its axis is a squircle: the point of the cross section plane belongs to an arm if its coordinates in that plane $\{\tilde x,z\}$ satisfy 
\begin{equation}
  \left|\frac{\tilde x}{r_\mathrm{disk}}\right|^n + \left|\frac{z}{r_z}\right|^n \le 1.
\end{equation}
Here $r_\mathrm{disk}$ and $r_z$ are two semi-axes of the squircle. The continuous parameter $n$ controls the shape of the cross section. If $n=2$ the squircle reduces to an ellipse, while larger $n$ makes it more similar to a rectangle.

The magnetic field of an arm is directed along the arm and varies in inverse proportion to the arm cross section as to guarantee its divergence-free structure. It is parameterized by the value $B$ at the intersection of the arm axis and the line passing through the Sun and the Galactic center. To minimize the number of fitting parameters we take the field to be constant over the arm cross section. 

In total each arm is described by 8 parameters: $x_0$, $y_0$, $b$, $\alpha$, $B$, $r_\mathrm{disk}$, $r_z$, $n$. The pitch angles $\alpha$ are taken the same for all arms and are controlled by the common pitch $\alpha_0$ which is allowed to vary freely while fitting. The $y$-coordinates of the arm origins are zero $y_0 = 0$ for all arms. To the contrary, the $x$-coordinates $x_0$ are individual for each arm and are set by hand to make our magnetic arms better match with the spiral design of the Galaxy. The adjustments of $x_0$ do not affect the result as long as $x_0$ is small compared to the distance from the Sun to the Galactic center. Indeed, in this case the change in $x_0$ is degenerate with the initial phase of the spiral arm $b$ which is a fitting parameter. We make use of this fact to roughly connect major Perseus and Scutum arms with the ends of the Galactic bar. The widths of the arms in the Galactic plane are also set to predefined values close to 1~kpc, see Table~\ref{tab:best_fit_pars}.

After these adjustments we are left with 4 fitting parameters per arm. For the nearby Local arm and the Sagittarius-Carina arm we fit all 4 parameters, while for the more distant Scutum arm only the strength $B$.

The Perseus arm is a special case. It contributes to the observables in the direction of the outer Galaxy at low Galactic latitudes where the Fan Region is located, which is one of the most prominent features on the synchrotron polarization maps. In order to fit the Fan Region we had to assume that the inner part of the Perseus arm close to the disk has stronger magnetic field as shown in Fig.~\ref{fig:GMF_xz}. We assumed for simplicity that the field takes two different constant values across the inner and outer sections of the arm (this can be viewed, of course, as a crude approximation for the field decreasing away from the disk). In total, there are three fitting parameters for the Perseus arm: the field strength $B$, the additional field  in the inner part $B^\prime$ (so that the total field in this part is $B+B'$) and the width of the arm $r_\mathrm{z}$. Note, that the direction of the field in the Perseus arm was fixed to be the same as in the Local arm. The width of the inner part $r_\mathrm{z}^\prime$ is fixed to 0.4~kpc. The phase $b$ and shift $x_0$ are chosen in such a way as to make the Perseus and Scutum arms symmetric with respect to the Galactic $z$-axis: $x_0(\mathrm{Perseus}) = -x_0(\mathrm{Scutum})$, $b(\mathrm{Perseus}) = b(\mathrm{Scutum}) + 180^\circ$, see Fig.~\ref{fig:GMF_xz} and Table~\ref{tab:best_fit_pars}.

\paragraph{Toroidal halo.} The field in the thick disk is not sufficient to account for the observed Faraday rotations --- a component extending  further in the Galactic halo is needed. Previous models have shown the necessity for a toroidal component above and below the Galactic disk with the azimuthal field directions opposite in the North and South. Such a field may arise as a result of the Galactic dynamo \citep{Wielebinski1993449}. This component is schematically shown in Fig.~\ref{fig:GMF_xz}.

It has been first noted in \citet{1997A&A...322...98H, Han:1999vi} and later confirmed by \citet{Pshirkov:2011um} that the combination of the thick disk and the toroidal halo reproduces well the characteristic all-sky $-+-+/-+$ North/South pattern of the Faraday rotations, provided the directions of the disk field around the Sun location and the toroidal component are opposite in the North and coincident in the South. As a result, the magnetic field in the Galactic Northern sky is somewhat weaker than in the South. If only Faraday rotation data are included, these two components give a reasonable fit to the data \citep{Pshirkov:2011um}. 

We model the toroidal components as cylindrical sections whose axes coincide with the Galactic axis, filled with purely azimuthal magnetic field. Each cylinder is described by 4 parameters: the outer radius $r_\mathrm{out}$, lower bound $z_\mathrm{min}$, upper bound $z_\mathrm{max}$ and the strength of the field $B_\mathrm{tor}$.

\paragraph{X-shape field.} When the synchrotron polarization data are included, we could not fit both RM and synchrotron maps with the thick disk and toroidal halo. With only these two components one cannot reproduce the change of the sign of the Q-parameter around the meridian line $l=0$, which is particularly visible in the data in the Northern hemisphere, see Fig.~\ref{fig:data}. As was first noted in \citet{JF_GMF_1}, this can be achieved by adding a second halo component --- the X-shape field --- in the inner part of the Galaxy. Topologically this field corresponds to the Model C of \citet{Ferriere1}. Similar fields are observed in some other galaxies viewed edge-on \citep{Beck:2019jyi}. To reduce the number of parameters, we take the X-field lines to be symmetric with respect to the Galactic plane, axially symmetric and having everywhere the same inclination angle $\theta$ with respect to the Galactic axis. The footprint of the X-field in the disk extends up to $r_X \sim 6$~kpc from the center. The X-field is schematically shown in Fig.\ref{fig:GMF_xz}. 

The fitting parameters of the X-field are the strength $B_X$ at $z=0$, the radius $r_X$ of the X-field footprint at $z=0$ and the inclination angle $\theta$. The radial profile of the X-field was chosen to be constant. Thus, the footprint of the X-field on a plane above or below the Galactic plane is a ring of inner radius $|z|/\cos\theta$ and outer radius $|z|/\cos\theta + r_X$. The area of this ring grows with $|z|$; the strength of the field decreases in inverse proportion so as to keep the field divergence-free.

\paragraph{Local Bubble.} The three components described so far reproduce well the overall pattern of both RM and synchrotron data. However, when normalized to the rotation measures, they underestimate the overall strength of the synchrotron polarization signal at high latitudes by about factor 2. In the previous models of \citet{JF_GMF_1,Unger:2023lob} where the synchrotron data were included in the fit this problem was solved by introducing the 'striation' which allows one to boost the synchrotron polarization signal relative to the rotation measures. We found that inclusion of the contribution from the Local Bubble may solve the problem without the striation factor. 
\begin{figure}
    \centering
    \includegraphics[width=0.85\linewidth]{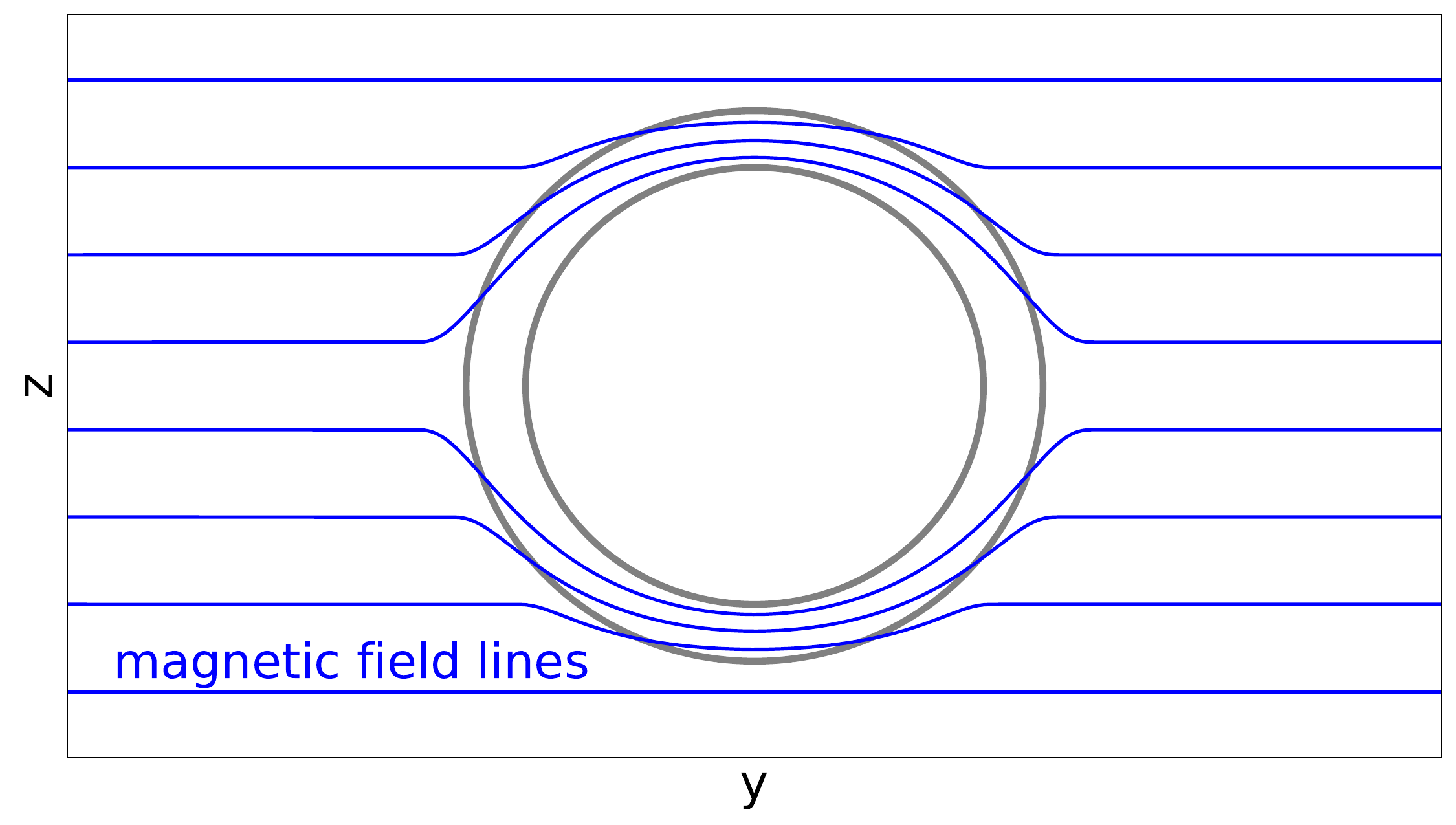}
    \caption{Schematic drawing of the Local Bubble and surrounding magnetic field lines. The radius of the bubble in our model is fixed to 200~pc, while the best-fit thickness of its wall was found to be 30~pc.}
    \label{fig:LB_artist}
\end{figure}

The Local Bubble is a cavity in the interstellar medium~\citep{2022A&A...661A.147L} created, presumably, by several local supernova. It is surrounded by clouds of cold dust compressed by hot gas resulted from supernova explosion(s). It has an irregular shape, but in the zeroth approximation can be fitted by an ellipsoid with half-axes of the size of $x=100$~pc by $y=200$~pc by $z=300$~pc \citep{LocalBubble_1, LocalBubble_2}. The Solar system is located in the inner part of the bubble relatively close to its center. 

It turns out that even in a crude approximation of a spherical bubble, the compressed magnetic field on the bubble walls can explain the missing part of the polarization intensity while having approximately right sky distribution and not spoiling the RM signal. The latter can even be improved by adjusting the parameters of the bubble as discussed in more detail below in Sect.\ref{sec:LocalBubble}. This is possible because the polarization parameters $Q$ and $U$ are quadratic in the field strength, while the RMs are linear. 

In our model the bubble is a spherical shell with the inner radius fixed to $r_\mathrm{LB} = 200$~pc. While fitting we tuned the thickness of the bubble wall $\delta r_\mathrm{LB}$, the coordinates of the center of the bubble $x_\mathrm{LB}$, $y_\mathrm{LB}$, $z_\mathrm{LB}$, and the direction $l_\mathrm{LB}$, $b_\mathrm{LB}$ of the magnetic field before compression. The field strength in the bubble wall is determined as a result of the compression of magnetic lines of the initially uniform field by the expanding spherically symmetric bubble. We assume that the field outside of the bubble is unperturbed, the field inside is zero, and the field in the wall is tangential to its surface and uniformly distributed across the wall in the radial direction. We also assume that it is symmetric with respect to the axis passing through the center of the bubble and parallel to the original uncompressed field. The field configuration is schematically shown in Fig.~\ref{fig:LB_artist}. Strictly speaking, such field is not divergence-free. However, by adjusting the dependence of the field strength on the distance from the axis one can make it divergence-free at scales larger than the thickness of the wall. The conservation of the magnetic flux at these scales implies the following relation:
\begin{equation}
    B_\mathrm{LB}(\theta) = B_\mathrm{LB,0} \left[1 + \frac{r_\mathrm{LB}^2}{2 r_\mathrm{LB} \delta r_\mathrm{LB} + \delta r_\mathrm{LB}^2}\right] \sin\theta,
\end{equation}
where $\theta$ is the angle between the direction of the initial field and the radius vector from the center of the bubble to the point in which the field is calculated. The strength of the initial field $B_\mathrm{LB,0}$ is taken to be the same as in the Local arm.

\subsection{Thermal electrons}
\label{sec:therm_elec}
The calculation of the RMs requires the knowledge of the distribution of free electrons in the Galaxy. In our study we adopt the simple plane-parallel model of the density of free electrons $n_e$:
\begin{equation}
    n_e(z) = n_0\,e^{-|z|/z_0}
\end{equation}
where ${n_0 = 0.015}$~cm$^{-3}$ is the mid-plane density and ${z_0 = 1.57}$~kpc is the scale height. This corresponds to the best-fit model of \citet{Ocker:2020tnt} where it was shown to correctly predict the dispersion measures (DM) of pulsars located above the Galactic plane with ${|b| > 20^\circ}$.

Similar values of $n_0$ and $z_0$ are used in more sophisticated models such as NE2001~\citep{2002astro.ph..7156C} and YMW16~\citep{2017ApJ...835...29Y}. Additionally, these models take into account the increase of the density of electrons in the spiral arms whose effect is most pronounced near the Galactic plane $|b| < 10^\circ$. Since in our analysis we treat the positions of magnetic arms as free parameters, the fixed positions of spiral arms of free electrons may significantly affect the results. For this reason we exclude the Galactic plane $|b| < 10^\circ$ from the analysis of RMs, as has already been stated in Sect.~\ref{subsec:RM}. At high Galactic latitudes the predictions of NE2001, YMW16 and plane-parallel models are approximately the same \citep{Ocker:2020tnt} and thus the use of the latter model does not lead to a loss of accuracy. At intermediate latitudes $10^\circ < |b| < 20^\circ$ the enhancement of thermal electron number density in two nearby spiral gaseous arms may be important and can affect our results. This however should not change the results significantly since the errorbars for the RM bins in this latitude bands are large.

In addition, we take into account the modifications of the thermal electron density caused by the Local Bubble. On the wall of the Local Bubble the density is higher and depends on its thickness $\delta r_\mathrm{LB}$ as
\begin{equation}
    n_\mathrm{LB} = n_0\,\frac{(r_\mathrm{LB} + \delta r_\mathrm{LB})^3}{(r_\mathrm{LB} + \delta r_\mathrm{LB})^3 - r_\mathrm{LB}^3}. 
\end{equation}
This expression follows from the assumption that the explosions which formed the Local Bubble swept the ambient electrons of original constant density $n_0$ to its walls. The value of the electron density inside the bubble is not important because there we assume the magnetic field to be zero.

\subsection{Cosmic ray electrons}
\label{sec:cr_elec}
Cosmic ray electrons (CRE)\footnote{Throughout the paper by CRE we mean both relativistic electrons and positrons.} are responsible for the Galactic synchrotron emission and are believed to be mostly accelerated by Galactic supernovae. Their propagation in the Galaxy \citep{Moskalenko:1997gh, Orlando:2013ysa} is of a diffusive nature and can be described by combined diffusion in the regular and turbulent magnetic fields.

In a simplified picture, CRE propagation can be describe as an isotropic diffusion determined by the single diffusion coefficient $D_0$. We modeled the distribution of the CRE in the Galaxy using the numerical code \texttt{DRAGON} \citep{Evoli:2016xgn}, taking into account CRE energy losses by the inverse Compton scattering and the synchrotron emission. As a template for synchrotron losses we used the modified version of \citet{JF_GMF_2} turbulent magnetic field with the normalization of all arms set to 5~$\mu$G.

An isotropic diffusion is assumed to be within the simulation volume of radius ${R_\mathrm{CRE} = 16}$~kpc and the halo height of ${H = 4}$~kpc. We neglected anisotropic diffusion and assumed also that the diffusion coefficient is spatially independent. The dependence on the CRE rigidity $R$ was taken to be the following: 
\begin{equation}
    D(R) = D_0\,\left(\frac{R}{R_0}\right)^\delta,
\end{equation}
where $D_0 = 3.6\times 10^{28}$~cm$^2$/s is the diffusion coefficient at rigidity $R_0=4$~GV and $\delta=0.47$ is the spectral index. The propagation parameters $H$, $D_0$ and $\delta$ were chosen based on the analysis of secondary-to-primary ratios by \citet{DelaTorreLuque:2024ozf}. Similar values were found in other studies, see \citet{Korsmeier:2021brc}. The sources of cosmic rays were assumed to be distributed in the Galaxy according to \citet{Lorimer:2006qs}. The dynamic range of the simulation goes from 1~GeV up to 1~TeV. Our resulting CRE distribution is the sum of the distribution of primary electrons and secondary electrons and positrons produced in interactions of cosmic ray protons. Note that the contribution of secondary leptons to the total CRE density does not exceed $10\%$ at energies of interest. Finally,  we normalize the distribution to the local measurements by the Alpha Magnetic Spectrometer (AMS-02) experiment \citep{PhysRevLett.122.101101}, as shown in Fig.~\ref{fig:CRE_AMS}. 

Modelling the Fan Region required an additional CRE component in the Perseus arm compared to the \texttt{DRAGON} output. Denoting as $L$ the length of the arm axis starting from its origin we increase the CRE density in the Perseus arm according to the bump-like profile:
\begin{equation}
    n(L) = \frac{f\,n_0}{(1 + e^{-k(L - L_1)})(1 + e^{-k(L_2 - L)})}
\end{equation}
where $n_0$ is the CRE density on the Solar vicinity, $f=0.7$ is the relative amplitude of the additional bump, the parameter ${k=2}$~kpc controls the sharpness of the bump boundaries, and $L_1$ and $L_2$ determine their coordinates. As a result, the CRE density is enhanced roughly in the direction from $l\simeq 110^\circ$ to $l\simeq 180^\circ$. The parameters $L_{1,2}$ were fixed from the geometry of the Fan Region. The amplitude of the enhancement was also fixed; its value in the fit is largely degenerate with the amplitude of the magnetic field. 

To the contrary, we do not consider the modification of the CRE density on the wall of the Local Bubble. Studying propagation of cosmic rays with full account of the details of the local magnetic field constitutes a separate problem and is beyond the scope of this paper.

Ideally, the CRE propagation should be fitted together with the GMF and include the effects of anisotropic diffusion. In practice, diffusion of CRE in space-dependent realistic GMF is a difficult problem which has not been solved so far \citet{Giacinti:2014xya, Kachelriess:2019oqu}. In this sense the additional CRE component required for the Fan Region in our model may indicate the necessity for deviation from the simplified isotropic picture.

Given the space and energy distributions of CRE in the Galaxy we calculate synchrotron Stokes $Q$ and $U$ parameters starting from the synchrotron emission of a single electron \citep{Landau:1975pou}. We thus do not adopt the commonly used power-law approximation of the electron spectrum for the volume emissivity and instead integrate numerically over the CRE spectrum. This allow us to take into account the curvature of the CRE spectrum as well as it's variation over the line of sight.

\section{Results of the global fit}
\label{sec:global_fit}
\begin{figure*}
    \centering
    \includegraphics[width=0.99\linewidth]{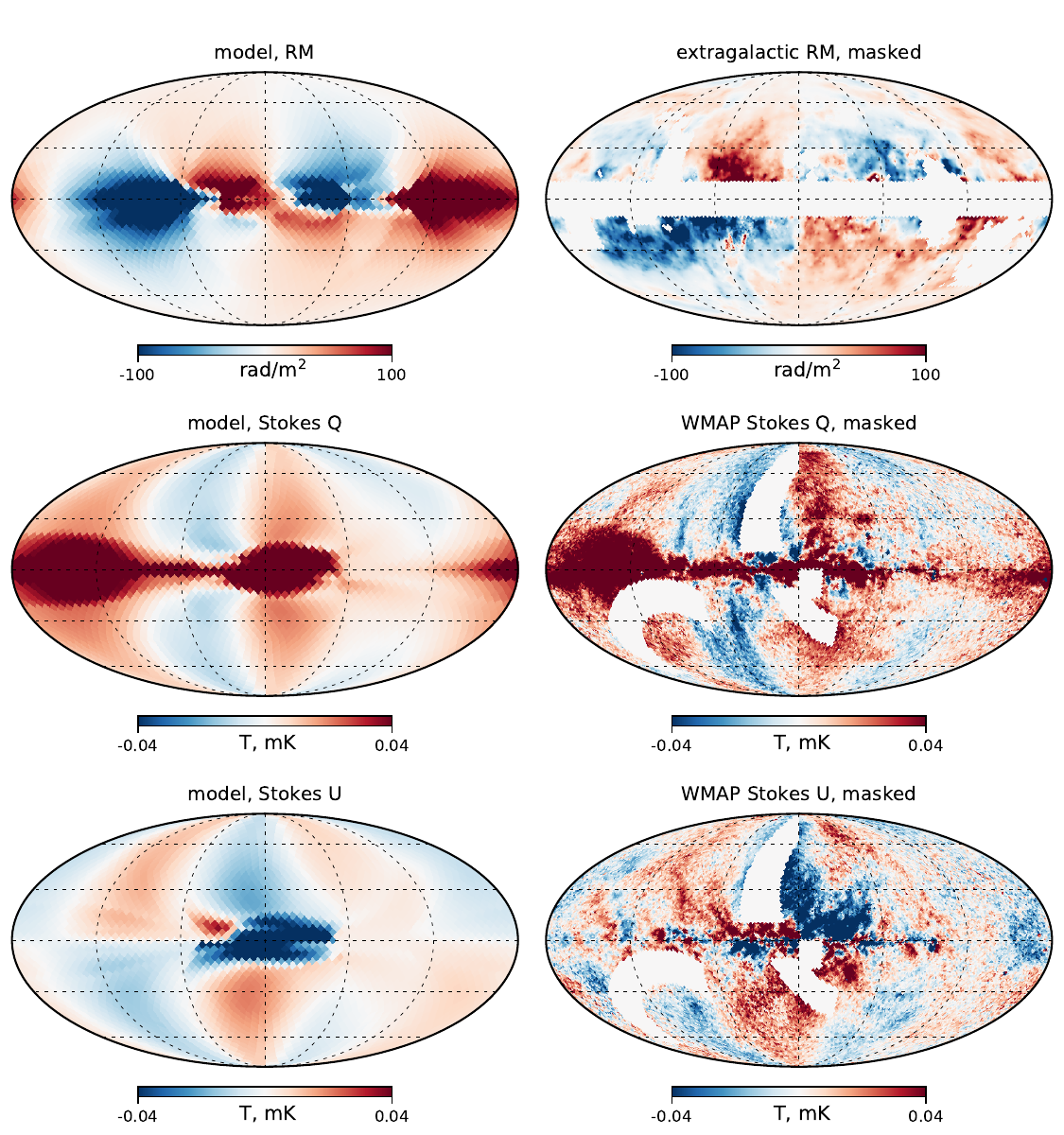}
    \caption{RM (top row) and polarized synchrotron skymaps for Stokes Q (middle row) and Stokes U (bottom row) produced with the best-fit model (left column) in comparison with the data (right column). Masks on the data maps are discussed in the Sect.~\ref{subsec:RM}.}
    \label{fig:skymaps}
\end{figure*}
The fitting of the model is performed by minimizing the $\chi^2$. We first adjusted individual GMF components, as well as their various combinations until we reproduced qualitatively the key features of the data, and only then performed the combined fit. The multi-parametric optimization was performed with the Nelder–Mead minimization method. To test whether the found minimum is a global one, we repeated the minimization with several different sets of initial parameters and varied the fitting step sizes. In all cases, the fit converged to the same parameter values, increasing the reliability of the results. After the convergence of the fit, we estimated the uncertainty of each parameter based on the curvature of the fit at the minimum \citep{10.5555/1403886}. Also, during the calculation of uncertainties, we ensured that the best-fit parameters correspond to the minimum of $\chi^2$ by computing the derivatives of $\chi^2$ with respect to each parameter and verifying that they are close to zero.

The model is fitted to the binned data as described in Sect.~\ref{subsec:bins_and_errors} with the binning scheme shown in Fig.~\ref{fig:binning}. We used only those bins, which are left after removing outliers and masking. All the data bins and corresponding errorbars used for fitting are shown with black points in Figs.~\ref{fig:bins_rm},\ref{fig:bins_stokes_q}, and \ref{fig:bins_stokes_u}. For each data bin we evaluated the model up to the distance of 30~kpc from the position of the Sun in the direction of the line of sight (LOS) passing through the center of the bin. We checked that if we take 10 evenly distributed LOSes per bin, the resulting $\chi^2$ changes very slightly, by 1-2\%. Therefore, for the sake of saving computational time during fitting, we used only the central LOS.

The results of the global fit are presented in Fig.~\ref{fig:skymaps} which shows the comparison between the model (left column) and the masked data (right column) for three quantities: the RMs and the $Q$ and $U$ polarization parameters. The polarized intensity map is shown in Fig.~\ref{fig:PI_skymaps}. One may observe an overall qualitative agreement between the model and data maps. The best-fit total $\chi^2/ndf$ is 1.36. The breakup of $\chi^2$ for the corresponding three contributions is given in Table~\ref{tab:chi2-breakup}, first three columns. The last two columns show what these quantities would be if the errors were calculated as mere variances in the bins, after removal of outliers. The over-estimation of errors in that case is obvious. The bin-by-bin comparison between the model and the data is given in the Appendix~\ref{appendix:comparison}. The maps of residuals are shown in Fig.~\ref{fig:residuals}. 
\begin{table}[ht]
    \caption{The result of the global fit.}
    \centering
    \begin{tabular}{cccc|cc}
        \hline\hline
        & $\chi^2$ & $\chi^2$/ndf & ndf & $\chi_\mathrm{var}^2$ & $\chi_\mathrm{var}^2$/ndf \\ \hline
        RM    & 544  & {\bf 1.92}  & 283  & 145 & 0.51 \\
        Q     & 385  & {\bf 1.11}  & 348  & 238 & 0.68 \\
        U     & 482  & {\bf 1.38}  & 348  & 251 & 0.72 \\ \hline 
        total & 1411 & {\bf 1.36}  & 1037 & 634 & 0.61
    \end{tabular}
    \tablefoot{First three columns show the performance of the best-fit model: the overall $\chi^2$ and $\chi^2$ per degree of freedom, as well as their breakup between RM and polarization Q and U maps. The last two columns show the same quantities when the standard deviation within the bins (calculated after removal of outliers) is taken as an error estimate.}
    \label{tab:chi2-breakup}
\end{table}

The best-fit parameters of the model are summarized in Table~\ref{tab:best_fit_pars}. The strength of the magnetic field in the Local Arm was found to be 3.5~$\mu$G which is higher than previously thought. The increased strength of the field is a consequence of making the {\em shape} of the cross section of the Local Arm a free parameter (the squircle parameter $n$). In previous models a constant height of the disk across the arms was assumed. To the contrary, in our model the best-fit value is $n=1.45$ causing the Local magnetic arm to be thicker in the center than above the Sun, see Fig.~\ref{fig:GMF_xz}. This shape of the cross section significantly improves the fit allowing for stronger field in the direction of the outer Galaxy. As a result, the model better describes the wave-like pattern in the RM maps of the outer Galaxy between $l \sim 90^\circ$ and $l\sim 270^\circ$, Fig.~\ref{fig:bins_rm}.

In the neighboring Sagittarius-Carina arm in the direction of the Galactic center the magnetic field was found to be about 1.3~$\mu$G, which is weaker than the field in the Local Arm. In addition, its direction is opposite to that of the field in the Local Arm. Reversing the field direction in the Sagittarius-Carina arm (so that it aligns with the direction of the field in the Local Arm) while keeping the same strength worsens the RM $\chi^2$ by $\Delta \chi^2 \approx 110$. Thus, we confirm the field reversal in accordance with \citet{reversal} and \citet{Pshirkov:2011um}. 

Similarly, we confirm the necessity of the X-field. Setting it to zero increases the total chi-square by $\Delta\chi^2 \approx 260$. The best-fit strength of the X-field near the Galactic plane was found to be 2 $\mu$G which is twice as large as in the \texttt{base} model of \citet{Unger:2023lob}. Our model prefers a sharp drop of the X-field strength at the Galactic radius of about 6~kpc. We discuss below in Sect.~\ref{sec:X-and-CR-escape} how much it costs in terms of $\chi^2$ to extend the X-type field to the solar vicinity as suggested by the arguments based on the escape of Galactic cosmic rays. 

The toroidal halo field is one of the key components of the model. If one sets the field in the toroidal halo to zero, the total chi-square nearly doubles: $\Delta\chi \approx 1010$. In the best-fit configuration the radial extensions of the norther and southern tori were found to be different, with the southern torus being larger than the northern one. However, the significance of this difference is not strong. When both tori are forced to have the same radius, the best-fit value of this common radius is $r_\mathrm{max} = 13.5$~kpc, which worsens the fit by only $\Delta\chi^2 \approx 20$.

Inclusion of the Local Bubble is a distinctive and novel feature of our model. As explained in Sect.~\ref{sec:LocalBubble}, we approximated it as a spherical shell of finite thickness with a compressed magnetic field on its wall. Notably, the fitted parameters of the Local Bubble were found to be close to those known from other observations. The best-fit thickness of the Local Bubble wall in our model is $\delta r_\mathrm{LB} = 30$~pc which is in perfect agreement with the typical thickness of $35$~pc determined in the recent analysis of \citet{2024arXiv240304961O} based on the distribution of the local dust. Given that the radius of the bubble was fixed to $200$~pc the field strength in the most compressed regions of the wall reaches $\sim 10$--$14$~$\mu$G in accordance with \citet{2006ApJ...640L..51A}, \citet{2019ApJ...873...87M} and \citet{2023A&A...673A.101K}. To be more precise, using the Chandrasekhar-Fermi method \citet{2006ApJ...640L..51A} found the plane-of-the-sky strength of the magnetic field $B_\perp = 8^{+5}_{-3}$~$\mu$G in the wall of the Local Bubble toward $(l, b) \approx (300^\circ, 0^\circ)$. This is in good agreement with the $B_\perp = 11$~$\mu$G predicted by our model in the same direction.  

The best-fit direction of the bubble field before compression in our model is $(l_\mathrm{LB}$, $b_\mathrm{LB}) = (230^\circ$, $-2^\circ)$, which is nearly opposite to the field of the Local Arm which has $(l, b) = (70^\circ, 0^\circ)$. Remarkably, \citet{LocalBubble_2} found that the direction of the pre-compressed bubble field lies in the range $(l, b)\approx (72^\circ \pm 1^\circ, 15^\circ\pm 2^\circ)$ by analyzing the polarized emission of the dust in the Galactic polar caps with $|b| > 60^\circ$. Similar initial direction was found in a more recent analysis of the dust emission by \citet{2024arXiv241017341O}. In view of the sign ambiguity inherent in the synchrotron polarization data, this is in perfect agreement with our fit. In fact, in our fit the reversed (as compared to the Local Arm) direction is preferred only by the RM data. It is worth noting, however, that the Local Bubble contribution to the RM strongly depends on the shape of the bubble wall, so the problem of the opposite field orientation should be re-examined together with a more accurate modelling of the bubble. 

The best-fit position of the center of the Local Bubble was found to be  shifted by $\sim 100$~pc from the position of the Solar System mostly in the direction of the positive $y$-axis. This is also generally consistent with the observations of the dust \citep{LocalBubble_2,2024arXiv240304961O}. Overall, the contribution from the Local Bubble is strongly favored by the data. Its removal worsens the fit by $\Delta\chi^2 \approx 660$. 

The shape and overall brightness of the Fan Region in our model generally matches the observations, as one can see from Fig.~\ref{fig:skymaps}. The main contributions come from the Local Arm and the Perseus Arm. The parameters of the former are strongly constrained by the RM data. Adjusting mainly the parameters of the Perseus Arm allowed us to fit the Fan Region. The correct vertical extension of the Fan Region was achieved by fitting the (outer) vertical height of the Perseus Arm which converged to $r_z = 1.2$~kpc. The field strength in the inner part around the Galactic plane was found to be $5.5$~$\mu$G in order to reproduce the brightest part of the Fan Region at $|b| < 10^\circ$. In the outer part of the Perseus Arm the best-fit field strength is about $3.5\,\mu$G.

The common pitch angle of the spiral magnetic field arms was an independent parameter of the fit and converged to the value around $\alpha_0 = 20^\circ$. As one can see from Fig.~\ref{fig:spiral_arms} this is close to the pitch angle of spiral arm segments as inferred from the Gaia data~\citep{gaia}. Despite the Galactic plane being masked out in the RM data, the fit is still sensitive to the pitch angle. For the RM skymap the sensitivity is mostly driven by the two nearest spiral arms (Local arm and Sagittarius-Carina Arm) which give significant contribution to the RM signal even at $|b| > 20^\circ$. Additionally, in the polarization skymaps the signal from the Fan Region depends strongly on the pitch angle of the Local and Perseus arms, see Sect.~\ref{sec:stability}.

\begin{table}[ht]
    \caption{List of the parameters of the model.}
    \centering
    \begin{tabular}{c|llll}
    \hline \hline
        Component    & Parameter         &  Value               & Units  & Fit   \\ \hline
        Local Arm    & $B$               &  -3.5 $\pm$ 0.1      & $\mu$G & free  \\
                     & $b$               &  -2.2 $\pm$ 0.8      & deg    & free  \\
                     & $n$               &  1.45 $\pm$ 0.03     &        & free  \\
                     & $r_\mathrm{z}$    &  0.73 $\pm$ 0.06     & kpc    & free  \\
                     & $r_\mathrm{disk}$ &  1.0                 & kpc    &       \\
                     & $x_0$             &  -0.15               & kpc    &       \\ 
                     & $y_0$             &  0.0                 & kpc    &       \\ \hline 
        Sagittarius- & $B$               &  1.3 $\pm$ 0.2       & $\mu$G & free  \\
        Carina Arm   & $b$               &  -80.0 $\pm$ 0.9     & deg    & free  \\
                     & $n$               &  2.3 $\pm$ 0.15      &        & free  \\
                     & $r_\mathrm{z}$    &  1.0 $\pm$ 0.3       & kpc    & free  \\
                     & $r_\mathrm{disk}$ &  0.8                 & kpc    &       \\
                     & $x_0$             &  1.37                & kpc    &       \\ 
                     & $y_0$             &  0.0                 & kpc    &       \\ \hline 
        Perseus Arm  & $B$               &  -3.5 $\pm$ 0.3      & $\mu$G & free  \\
                     & $B^\prime$        &  -2.0 $\pm$ 0.5      & $\mu$G & free  \\
                     & $b$               &  46.0                & deg    &       \\
                     & $n$               &  2.0                 &        &       \\
                     & $r_\mathrm{z}$    &  1.2 $\pm$ 0.13      & kpc    & free  \\
                     & $r_\mathrm{z}^\prime$ &  0.4             & kpc    &       \\
                     & $r_\mathrm{disk}$ &  1.1                 & kpc    &       \\
                     & $x_0$             &  -1.0                & kpc    &       \\ 
                     & $y_0$             &  0.0                 & kpc    &       \\ \hline 
        Scutum Arm   & $B$               &  4.9 $\pm$ 0.4       & $\mu$G & free  \\
                     & $b$               &  -134.0              & deg    &       \\
                     & $n$               &  2.0                 &        &       \\
                     & $r_\mathrm{z}$    &  0.8                 & kpc    &       \\
                     & $r_\mathrm{disk}$ &  1.0                 & kpc    &       \\
                     & $x_0$             &  1.0                 & kpc    &       \\ 
                     & $y_0$             &  0.0                 & kpc    &       \\ \hline
        common pitch & $\alpha_0$        &  20.0 $\pm$ 0.2               & deg    & free  \\ \hline
        Northern     & $B$               &  3.2 $\pm$ 0.4       & $\mu$G & free  \\
        Torus        & $z_\mathrm{min}$  &  1.18 $\pm$ 0.07     & kpc    & free  \\
                     & $z_\mathrm{max}$  &  2.1 $\pm$ 0.2       & kpc    & free  \\
                     & $r_\mathrm{max}$  &  9.1 $\pm$ 0.2       & kpc    & free  \\ \hline 
        Southern     & $B$               &  -3.2 $\pm$ 0.7      & $\mu$G & free  \\
        Torus        & $z_\mathrm{min}$  &  -2.5 $\pm$ 0.4      & kpc    & free  \\
                     & $z_\mathrm{max}$  &  -1.22 $\pm$ 0.11    & kpc    & free  \\
                     & $r_\mathrm{max}$  &  14.0 $\pm$ 1.6      & kpc    & free  \\ \hline 
        X-field      & $B$               &  1.8 $\pm$ 0.2       & $\mu$G & free  \\
                     & $r_\mathrm{max}$  &  6.2 $\pm$ 0.2       & kpc    & free  \\
                     & $\theta$          &  28.0 $\pm$ 3.4      & deg    & free  \\ \hline
        Local        & $x_\mathrm{LB}$   &  -8.2                & kpc    &  \\
        Bubble       & $y_\mathrm{LB}$   &  0.095 $\pm$ 0.01    & kpc    & free  \\
                     & $z_\mathrm{LB}$   &  -0.05 $\pm$ 0.015   & kpc    & free  \\
                     & $r_\mathrm{LB}$   &  0.2                 & kpc    &  \\
                     & $\delta r_\mathrm{LB}$ &  0.03 $\pm$ 0.004  & kpc    & free  \\
                     & $l_\mathrm{LB}$   &  230.0 $\pm$ 1.6     & deg    & free  \\
                     & $b_\mathrm{LB}$   &  -2.0 $\pm$ 2.0      & deg    & free  \\ \hline
    \end{tabular}
    \tablefoot{See Sect.~\ref{sec:model} for the description of each parameter.}
    \label{tab:best_fit_pars}
\end{table}

\section{Discussion}
\label{sec:discussion}
\subsection{Local Bubble and nearby loops}
\label{sec:LocalBubble}
The Local Bubble introduced in our model significantly improves the fit for the polarization data and, to a lesser extent, for the RMs. Below we discuss in more detail its role in these fits. 

On the RM skymap the effect of the Local Bubble is clearly visible around the northern polar cap. In this region between $0^\circ \lesssim l \lesssim 180^\circ$ the Local Bubble gives a positive contribution to the RM which appears as a nearly uniform orange blur in Fig.~\ref{fig:skymaps}. In the binned representation one can see this area as three data bands with $50^\circ < b < 80^\circ$ with mostly positive RMs, see Fig.~\ref{fig:bins_rm}. This is the region of the RM sky that forces the field of the bubble to take direction opposite of the Local Arm. We should stress, however, that the sign and the amplitude of the bubble contribution to the RM depends significantly on the shape of its wall and the relative position of the solar system within the bubble. For instance, in our model where the bubble is spherical, this contribution can be set to zero by placing the solar system in the center as the field on the wall, which is purely tangential, is then perpendicular to the line of sight for all directions. It is clear, therefore, that a more detailed modelling of the bubble shape is required to firmly determine whether the direction of the field is parallel or anti-parallel to the field in the Local Arm.

The sign flip of the initial bubble field has no effect on the polarization maps. For both initial directions the Local Bubble produces the correct shape of the signal in $Q$ and $U$ maps near the Galactic poles, provided the bubble field is aligned to that in the Local Arm. The bubble contribution then almost completely compensates the deficit of the polarized intensity resulting from the other components when they are normalized to the RM data. It thus works similarly or even better than the 'striation' of the magnetic fields assumed in previous works. We think, therefore, that the alignment of the fields in the bubble and the Local Arm is a more robust result of our fit than the flipped direction of the bubble field which may be an artifact of our over-simplified modeling of the bubble geometry. 

\begin{figure}
    \centering
    \includegraphics[width=0.99\linewidth]{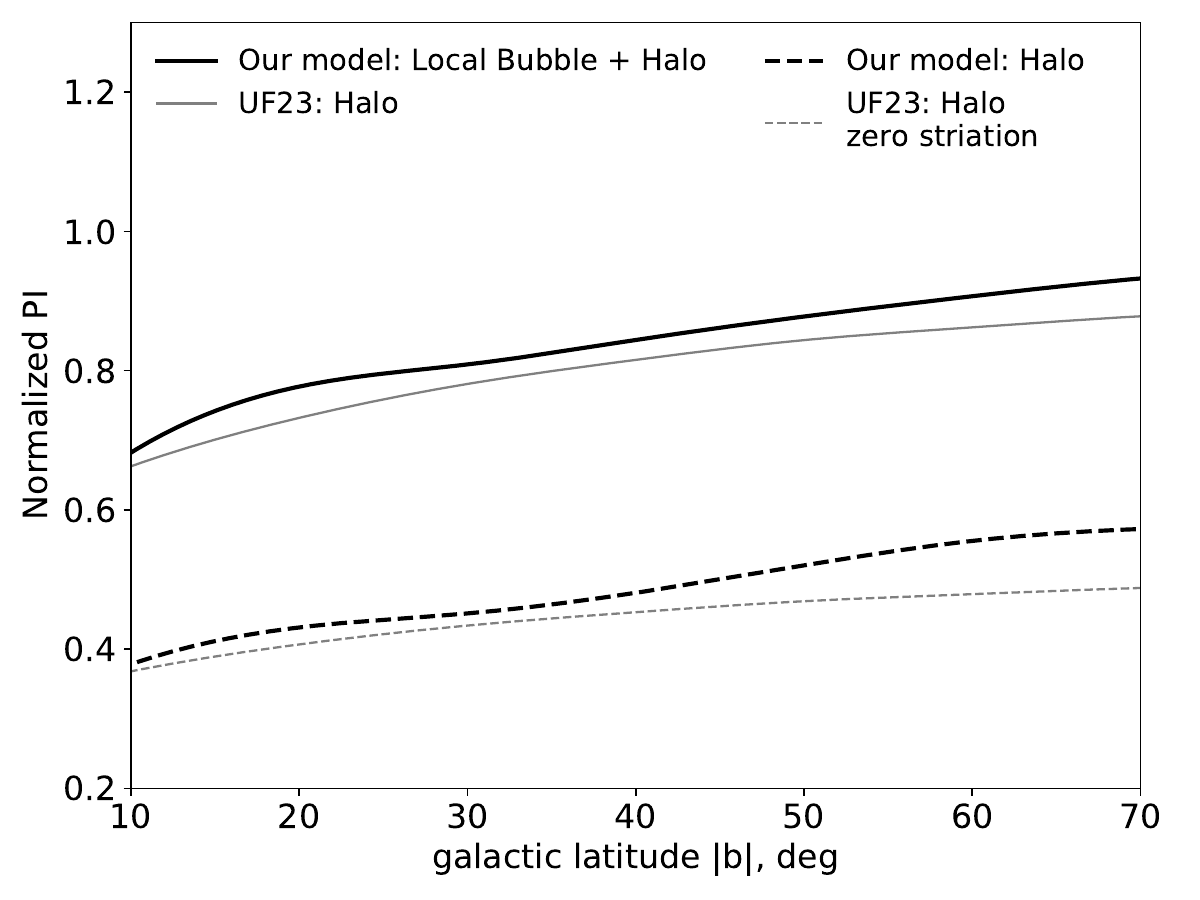}
    \caption{Polarized intensity of the halo and halo+Local Bubble above given latitude $b$ normalized to the polarized intensity of the full model in the same region of the sky. In this plot "Halo" means everything above $|z| > 1$~kpc. See text for the discussion. }
    \label{fig:PI_frac}
\end{figure}

\begin{figure}
    \centering
    \includegraphics[width=0.99\linewidth]{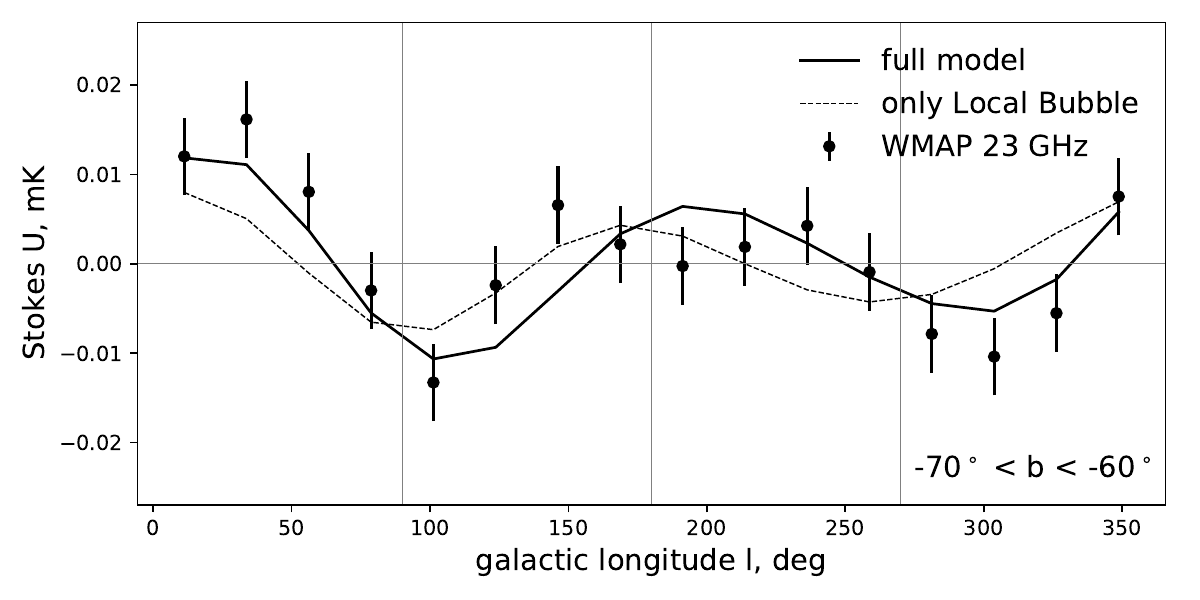}
    \caption{Profile of the Stokes U signal, calculated with the full model and the contribution from the Local Bubble.}
    \label{fig:locbub_u}
\end{figure}
The role of the Local Bubble in fitting the polarization data is illustrated in Fig.~\ref{fig:PI_frac} which shows the polarized intensity $\text{PI}=\sqrt{Q^2+U^2}$ integrated over the polar caps above given Galactic latitude $b$, as a function of $b$. The thick dashed line shows the fractional contribution of the "halo" part of the field defined here as everything at $|z|>1$~kpc, normalized to the full model. The thick solid line shows the contribution of the "halo" and the Local Bubble, without other "disk" fields at $|z|<1$~kpc. One can see that these two contributions together practically saturate the PI at high latitudes in roughly equal shares. For comparison, the same quantity is plotted in the model of \citet{Unger:2023lob} with (thin solid curve) and without (thin dashed curve) the striation. The effects of the Local Bubble and striation are quite similar. The contribution of the Local Bubble to the Stokes U is explicitly shown in the Fig.~\ref{fig:locbub_u}.

The significance of the Local Bubble can also be understood with the simple analytical estimate. The brightness of the PI is proportional to the square of the magnetic field $B^2$, the CR electron density $n_\mathrm{CRE}(z)$ and the distance $d$. At high Galactic latitudes in the model of \citet{Unger:2023lob} the main contribution to the PI comes from the the toroidal halo, which starts at $z \sim 1$~kpc and has a width of $d \sim 1$~kpc. Thus ${\mathrm{PI}_\mathrm{halo} \propto (3 \mu\mathrm{G})^2\,n_\mathrm{CRE}(1\,\mathrm{kpc}) \cdot \mathrm{kpc}}$. At the same time the contribution from the Local Bubble reads: ${\mathrm{PI}_\mathrm{LB} \sim (10 \mu\mathrm{G})^2\,n_\mathrm{CRE}(0)\cdot 0.03\,\mathrm{kpc}}$. Noting that ${n_\mathrm{CRE}(1\,\mathrm{kpc}) \approx n_\mathrm{CRE}(0)/2}$ one comes to the conclusion that both contributions are approximately equal:
\begin{equation}
    \mathrm{PI}_\mathrm{halo} \approx \mathrm{PI}_\mathrm{LB}.
\end{equation}

The Local Bubble also affects propagation of cosmic rays from nearby sources. In particular, it helps to explain the cosmic ray flux at the knee by a contribution from the Vela supernova remnant \citep{Bouyahiaoui:2018lew}. Also, interactions of cosmic rays with the gas on the walls of the Local bubble can give a significant contribution to the astrophysical neutrino flux \citep{Andersen:2017yyg,Bouyahiaoui:2020rkf}. 

There are other large-scale anomalies in the data which are also, presumably, due to supernova bubbles, but viewed from outside, such as Loop I, Loop II, Loop III, etc. The fact that these anomalies are prominent on the polarization sky suggests that the contribution of the Local Bubble is also sizeable. Inversely, it is possible that these anomalies can also be modelled in a similar way and included in the fit, rather than masked \citep{GMIMS_loopII}. We leave this interesting question for future work.

\subsection{Local X-shape field and escape of cosmic rays} 
\label{sec:X-and-CR-escape}
The ratio of secondary to primary fluxes of cosmic rays indicates that locally most of the cosmic rays escape from the Galaxy before interaction with the interstellar gas (for the review see \citet{Kachelriess:2019oqu} and references therein). In the simplified models of cosmic ray propagation the escape is described by diffusion. However, for pure turbulent field the required diffusion coefficient turns out to be too high, corresponding to unrealistically small magnetic field of the order of $10^{-5}$~$\mu$G. In \citet{Giacinti:2017dgt} it was shown that local $z$-component of the coherent GMF with the strength of a fraction of $\mu$G can resolve this cosmic ray escape problem. In the recent collection of GMF models of \citet{Unger:2023lob} the local $z$-component is present in most of the models (see Fig.~16 in \citet{Unger:2023lob}). 

In our best-fit model the X-field vanishes beyond the galactocentric radius of 6.2 kpc. Adding the local $z$-component beyond 6.2 kpc with the strength of $0.2$~$\mu$G slightly worsens the fit. If the direction of this additional component is the same as the direction of the X-field (i.e., towards positive $z$) the increase in the chi-square is $\Delta \chi^2 \approx 90$. For the opposite direction the worsening is smaller, $\Delta \chi^2 \approx 30$. In both cases all the increase comes from the RM fit since this field is too weak to contribute to the polarized intensity which is quadratic in the magnetic field strength. On the other hand, the contribution to the RM is of the same order as the effect of the Local Bubble discussed in Sect.~\ref{sec:LocalBubble}. Thus, the search for such a weak additional component can only be performed together or after the accurate modeling of the Local Bubble.

\subsection{Fan Region}
\label{sec:fan_region}
The Fan Region is a bright feature of the radio sky located near the Galactic plane roughly in the direction $90^\circ < l < 180^\circ$. It is visible as a red spot in the Stokes~Q meaning that it contains magnetic field mostly parallel to the Galactic plane.

The distance to the Fan Region remains uncertain. For a long time it was believed to be a local feature due to its exceptional brightness, high degree of polarization and bubble-like shape. In particular, \citet{2021ApJ...923...58W} reproduced the brightness and the shape of the Fan Region and Loop~I with a bundle of local, narrow magnetic filaments with a width of~$\sim$~20~pc possessing strong magnetic field of about $24$~$\mu$G.

However, the model of \citet{2021ApJ...923...58W} seems to be in tension with recent observations at lower frequencies. First, there are indications that the Loop-I is not a local feature but a Galactic-scale outflow located at 3-5~kpc from the Galactic center \citep{Zhang:2024mcd,Churazov:2024knj}. Second, \citet{fan} argued that at least 30\% of the polarized emission of the Fan Region at frequencies around 1 GHz or higher must come from a distance of 2 kpc or more. The conclusion of \citet{fan} was based on the observation that the portions of the 1.5~GHz Fan Region emission are depolarized by $\approx$ 30\% by distant ionized gas structures, indicating that a corresponding fraction of the emission originates from more than 2 kpc away. Additional evidence comes from the observations of the Radcliffe Wave located 0.5-1~kpc away in the direction to the Fan Region~\citep{2020Natur.578..237A}. Based on the orientation of the magnetic field of the Radcliffe Wave \citet{2024arXiv240603765P} concluded that it is unlikely that the Fan Region is at a distance $< 1$~kpc.

In this regard, our model is the first attempt to demonstrate that the Fan Region can be naturally incorporated as a part of the large scale coherent GMF, as suggested by~\citet{fan}. In contrast, in previous GMF models, the Fan Region was considered a local feature and was either masked out or ignored.

The two main ingredients that allowed us to fit the Fan Region are a larger (compared to previous models) pitch angle of the disk field and increased electron density in the Perseus arm. The former shifts the position of the peak intensity towards $l\approx 135^\circ-140^\circ$, while the latter allows to reach the required brightness. The larger pitch angle was already predicted by~\citet{fan}. Using simple geometrical analysis~\citet{fan} concluded that in order to fit the Fan Region into the large-scale GMF model the pitch angle of the magnetic field in the disk should be in the range 15$^\circ$-20$^\circ$. The pitch angle of our model converged to 20$^\circ$ thus confirming the conclusions of~\citet{fan}.

On the other hand, the increased CRE density in the Perseus arm may indicate either an efficient CRE transport along the arm due to the anisotropic diffusion or the presence of the local cosmic ray accelerators. This increase in our model was introduced artificially. More reliable distribution could be obtained by including the aforementioned effects into the CR propagation codes such as GALPROP~\citet{Strong:1998pw,Porter:2021tlr} or DRAGON~\citet{Evoli:2016xgn}. Note also that the exact distribution of the magnetic field along the LOS (in particular towards the Fan Region) cannot be reliably inferred without invoking the tomographical methods.

\subsection{Striated field and stability of the results}
\label{sec:stability}
The model presented in this study is quite different from the existing ones in several respects, notably it has two additional components (the Local Bubble and the Fan Region), the disk field pitch angle of 20$^\circ$ and zero striated fields. It is therefore instructive to check how these new features affect each other and the overall structure of the GMF model.

As a first test we included the striation coefficient $\xi$ as an additional free parameter of the fit. Following the definition from \citet{Unger:2023lob} the magnetic field strength used for the calculation of the synchrotron skymaps was rescaled as $B^\prime = B(1+\xi)$. Performing the fit with all the components of our model and this additional free parameter, the best-fit value of $\xi$ was found to be compatible with zero: $\xi = 0.01 \pm 0.03$. This is to be expected since the striation factor affects only the synchrotron intensity while the Local Bubble improves also the RM model, and thus the latter is preferred by the fit. On the other hand, the fit without the Local Bubble (with the magnetic field strength on the bubble wall set to zero) and without the additional CRE component in the Perseus arm resulted in $\xi \approx 0.3$ compatible with $\xi = 0.346$ in the \texttt{base} model of \cite{Unger:2023lob}, but yielded a poorer fit for the RMs.

Second, we explored the reason of larger pitch angle in our model. We found that the fit of only RM data prefers the pitch of about 13$^\circ$. However, the profile of the best fit $\chi^2$ as a function of the pitch angle is flat between 10$^\circ$ and 20$^\circ$: the best fit $\chi^2$ worsens by only $\Delta\chi^2 \approx$~20 if the pitch is fixed to 20$^\circ$ instead of 13$^\circ$. The same small pitch angle is found if the synchrotron data is included, but the Galactic plane and the Fan Region are masked out, in agreement with~\citet{Unger:2023lob}. The strong preference for the pitch angle of 20$^\circ$ is found after the inclusion of the Fan Region. Thus we conclude that the fit prefers larger pitch to fit the Fan Region. This worsens the RM fit, but only slightly.

\subsection{General remarks}
\label{sec:general_remarks}
Our Galaxy is believed to be a typical representative of the class of spiral galaxies. Therefore, it appears likely that the structure of its magnetic field should be similar to that of other spirals. Below we discuss how well our model fits into the bulk of knowledge about other galaxies.

Our fit requires the presence of the disk and halo fields in our Galaxy, which were also found in external spiral galaxies observed so far. We found that the pitch angle of the disk field is close to that of spiral gaseous arms. The rough similarity of the magnetic and stellar pitch angles were also observed in other spirals \citep{Beck:2019jyi}. 

From the polarization observations of other galaxies one finds that the strongest field of the disk often resides in interarm regions, while the regular field in the gaseous arms is somewhat suppressed. As it is clear from Fig.~\ref{fig:GMF_xz}, if our Galaxy was observed from outside, its spacial distribution of the polarized intensity would be the opposite: the brightest parts would rather coincide with the gaseous arms. In order to see whether this is an unavoidable feature of our model we replace the constant field in the arms by hollow magnetic profiles such that the magnetic field in the inner 70\% of the arm section is zero. Notably, this change of the profile does not spoil the RM fit while shifting the strongest polarized emission regions to the interarm space. The polarization skymaps also remain practically unchanged at latitudes above $\sim 30^\circ$ where the polarized emission is mostly produced by the Local Bubble and the magnetic tori. However, one should adjust the strength of the field in the arms to have the correct emission near the Galactic plane. We do not see any fundamental contradictions that would prevent this modification of the model. We leave the detailed analysis of this possibility for future work.

Another feature that is observed in some other galaxies is field reversals. As one can see from Fig.~\ref{fig:spiral_arms}, our model has one field reversal in the disk. The galaxies with one reversal in the disk have also been observed, see~\citet{Beck:2019jyi}.

\section{Conclusions} 
\label{sec:conclusions}

In this paper we presented the new model of the regular magnetic field in the Galactic halo, by which we understand the field everywhere except the thin disk. 
The model makes use of the most recent catalog of the Faraday rotation measures and the synchrotron polarization data by WMAP at 23 GHz. It builds on the previously existing models but, apart from using the most recent data, has several novel features as compared to previous works. 

For the first time we took into account the contribution of the magnetic field in the wall of the Local Bubble. By constructing an explicit, albeit rough, model of the Local Bubble as a sphere with the compressed magnetic field concentrated in the thick wall, we have shown that its contribution to the polarized intensity can explain the observations at high Galactic latitudes without contradiction with the RM data. This is in contrast with the previous works where the deficit of model polarized intensity in these regions had to be compensated by assuming the "striation" of the field. The best-fit parameters of the bubble were found to be in good qualitative agreement with the existing independent measurements.  

For the first time we have fitted the Fan Region --- the large bright region in the outer galaxy which was masked out in the previous works that were using the polarization data. There recently appeared arguments that a large part of the polarized emission in this region comes from distances of $1-2$~kpc and thus cannot be explained by a local anomaly. In order to fit the overall brightness and morphology of the polarized emission we had to assume that the magnetic field in the Perseus arm varies along the arm cross section, being stronger in the inner part close to the disk. 

With these new features included, the best-fit value of the pitch angle of the magnetic arms comes out close to $\sim 20^\circ$ which is about 2 times larger than found in previous works. This value, however, is in good agreement with the recent measurements by Gaia of the pitch angle of stellar arms neighbouring the solar system. 

Finally, we have developed a systematic method of error estimation which gives control over the relative contribution of the data of different nature (the RM and polarization data in our case) into the global fit. This method takes into account the errors of individual measurements, the statistics of measurements in the bins as well as the contribution of fluctuations at length scales exceeding the bin size. We found that the errors estimated by this method are typically smaller than were used in previous works, thus increasing the constraining power of the data.

\begin{acknowledgements}
We would like to thank Ioana Codrina Maris for illuminating discussions and Denis Allard for useful comments on the manuscript. Our special thanks to Michael Unger, Maxim Pshirkov and Vincent Pelgrims for thorough reading of the manuscript and numerous constructive comments. The work of AK and PT is supported by the IISN project No. 4.4501.18.
\end{acknowledgements}

\bibliographystyle{aa}
\bibliography{references}

\begin{appendix}
\section{Removal of outliers}
\label{appendix:cleaning}
Each data bin that we consider (a rectangular region on the sphere limited by lines of constant $l$ and $b$), we fit with a Gaussian. Namely, we assume that the measurements contained within the bin are samples from a Gaussian distribution with unknown parameters. By fitting data with a Gaussian we determine the mean $\mu$ and width $\sigma$ of this distribution, taking into account errors of individual measurements.

To do this we construct a special likelihood function $\mathcal{L}$. Denoting as $\mu_\mathrm{i}$ and $\sigma_\mathrm{i}$ the mean and error of each measurement in the bin it reads:
\begin{equation}
    \mathcal{L} = \prod_{i=1}^{N} \int\limits_{-\infty}^{\infty}\frac{1}{\sqrt{2\pi\sigma^2}\sqrt{2\pi\sigma_\mathrm{i}^2} } 
    \exp\left({-\frac{(\mu - x^\prime)^2}{2\sigma^2}}
    {-\frac{(\mu_\mathrm{i} - x^\prime)^2}{2\sigma_\mathrm{i}^2}}\right) \,\mathrm{d}x^\prime.
\end{equation}
where $N$ is the total number of measurements within the bin. Physically each integral under the product means the probability of observing the value $\mu_\mathrm{i}$ given the parameters of the distribution $\mu$ and $\sigma$ and an observational error $\sigma_\mathrm{i}$. This likelihood explicitly relies on the assumption of Gaussianity of the errors and in the limit  $\sigma_\mathrm{i} \to 0$ reduces to a standard likelihood function.
\begin{figure}
    \centering
    \includegraphics[width=0.95\linewidth]{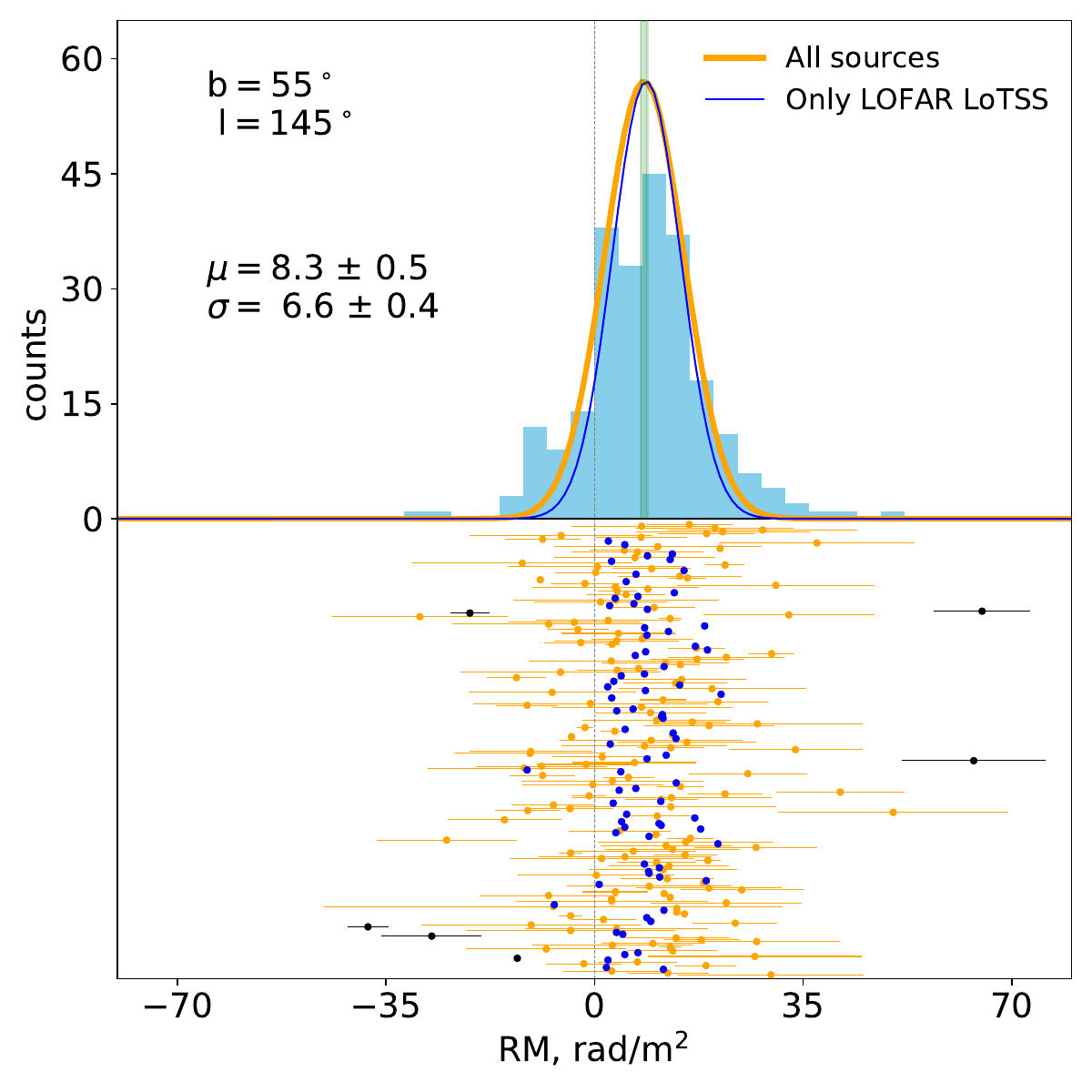}
    \caption{Example of RM bin fitted with the gaussian.  Points at the bottom of the figure indicate individual measurements within the bin together with their observational errors. Black points were discarded as outlier during the fit, blue points mark the LOFAR LoTSS data and orange points mark all other data. Blue boxes show the histogram of all sources except for outliers. Orange curve is the gaussian obtained as a result of the fit to all data (including LOFAR) and the pale green band is 68\% confidence interval of its $\mu$. The parameters of this gaussian are indicated on the plot. The blue curve is the best-fit gaussian if only LOFAR LoTSS data is used.}
    \label{fig:bin_gauss_fit}
\end{figure}

Evaluating the integral we find 
\begin{equation}
    \mathcal{L} = \prod_{i=1}^{N} \frac{1}{\sqrt{2\pi(\sigma^2 + \sigma_\mathrm{i}^2)}} \exp\left({-\frac{(\mu - \mu_\mathrm{i})^2}{2(\sigma^2 + \sigma_\mathrm{i}^2)}}\right).
\end{equation}
The best fit values of $\sigma$ and $\mu$ are obtained by minimizing the log likelihood $\chi^2 = -2\log\mathcal{L}$. When applied to the real data we use an iterative procedure which also allow us to self-consistently remove outliers. After fitting the bin we remove those measurements whose values deviate from the mean $\mu$ by more than $3\sqrt{\sigma^2+\sigma_\mathrm{i}^2}$ and then repeat the fitting. The algorithm terminates when no outliers are detected. As a final step we determine $1\sigma$ confidence intervals ${\mu \pm \sigma_\mu}$ and ${\sigma \pm \sigma_\sigma}$ by exploring the contour of ${\chi^2 = \chi^2_\mathrm{min} + 1}$. 

An example of the bin is shown in Fig~\ref{fig:bin_gauss_fit}. The best fit values as well as its uncertainties are shown in the legend. The resulting Gaussian curve is narrower than the histogram since it shows how the distribution would look like assuming zero measurement errors. In total the bin initially contained 245 sources, 8 of which were detected as outliers (only 6 are visible in the plot). Since the algorithm takes into account errors of individual measurements, it can be noticed that a source with an RM of about -30~rad/m$^2$ and a large error is included in the fit, while a source with a smaller RM of $\sim$~25~rad/m$^2$ and a much smaller error is marked as an outlier.
\begin{figure}
    \centering
    \includegraphics[width=0.95\linewidth]{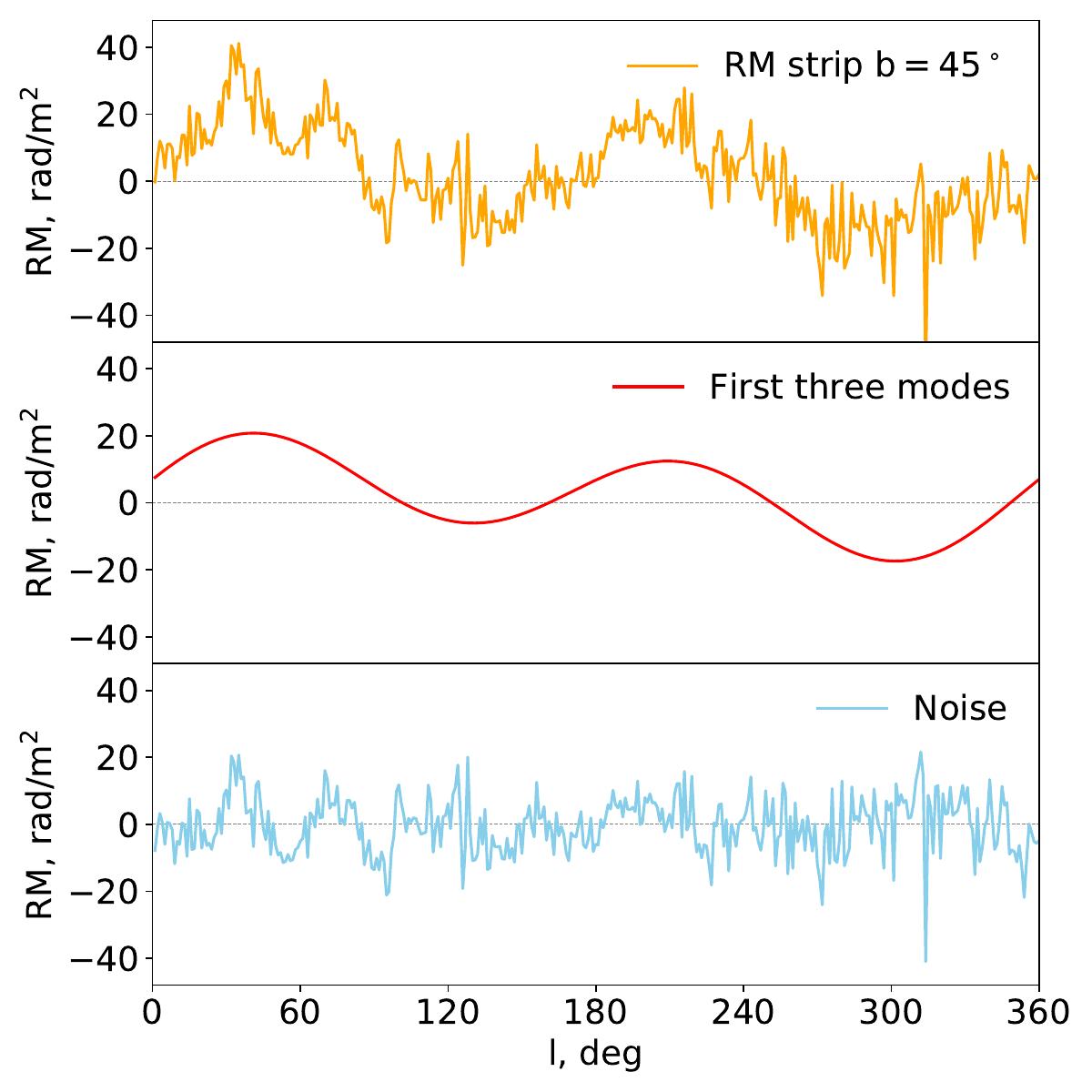}
    \caption{Fourier decomposition of the RM strip $40^\circ < b < 50^\circ$. The upper panel shows the RM signal binned into 360 bins by $l$. The middle panel shows the first three Fourier modes (namely, the constant and $360^\circ$ and $180^\circ$ wavelength modes) of the RM signal, while the lower panel shows the contribution from all other modes except for the first three.}
    \label{fig:strip_fourier}
\end{figure}

\section{Derivation of the bin variance}
\label{appendix:Derivation}
Here we derive the variances of the bins associated with the coherent fluctuations in the ISM. First we derive the formula and then discuss how it is applied to the data.

Assume that the observable $f$ depends on the discrete positions $x^i$ and denote $f^i = f(x^i)$. The positions are evenly spaced so that $x_i - x_{i - 1} = \Delta x$ for any $i$. Assume also that $f$ can be represented as a sum of the signal $f_0$ and a random noise $\epsilon$:
\begin{equation} \label{eq:n1}
    f^i = f^i_0 + \epsilon^i.
\end{equation}
Consider then the quantity $f_\mathrm{L}^i$ which is the value of $f$ averaged over $N$ nearby positions, or in other words averaged over the bin of the size $L = N\,\Delta x$
\begin{equation} \label{eq:n2}
    f_\mathrm{L}^i = \frac{1}{N}\sum\limits_{j=i}^{i+N-1} f^j.
\end{equation}
We are interested in its variance $\sigma_\mathrm{L}^2$:
\begin{equation} \label{eq:n3}
    \sigma_L^2 = \langle \left(f_\mathrm{L}^i - \langle f_\mathrm{L}^i \rangle \right)^2 \rangle.
\end{equation}

The angle brackets here denote averaging over different realizations of the noise $\epsilon$ while keeping the same signal $f_0$. We further assume that the statistical properties of the noise do not depend on the position $x^i$ and that the noise has zero mean $\langle \epsilon \rangle = 0$. Substituting \ref{eq:n1} and \ref{eq:n2} into \ref{eq:n3} one gets
\begin{equation}\label{eq:n4}
    \sigma_L^2 = \langle \left( \frac{1}{N}\sum\limits_{j=i}^{i+N-1} \epsilon^j \right)^2 \rangle = \frac{1}{N^2}\sum\limits_{j=0}^{N-1} \sum\limits_{q=0}^{N-1} \langle \epsilon^j \epsilon^q \rangle.
\end{equation}
As we assumed that the noise does not depend on $x$, the correlator $\langle \epsilon^j \epsilon^q \rangle$ is the function only of the difference $|x_j-x_q|$ but not the position itself. This allows one to evaluate the double sum in the Eq.~\ref{eq:n4} by collecting and counting equal terms. Denoting the correlator $\langle \epsilon^j \epsilon^q \rangle = r(|x_j-x_q|)$ one obtains the following expression:
\begin{equation}\label{eq:n5}
    \sigma_L^2 = \frac{1}{N} \sum\limits_{j=-N+1}^{N-1} \left(1-\frac{|j|}{N}\right)\,r(|j\,\Delta x|),
\end{equation}
where $r(x)$ is the correlation function of the noise. If the noise can be extracted from the measurements, this formula allows one to directly calculate the variance for the extended bin of the size~$L$. 

In order to clarify the physical meaning of \ref{eq:n5} it is useful to consider the continuous limit $\Delta x \to 0$ in which the sum in \ref{eq:n5} can be replaced with an integral:
\begin{equation}\label{eq:n6}
    \sigma_L^2 = \frac{2}{L} \int\limits_{0}^{L} \mathrm{d}x \left(1-\frac{x}{L}\right)\,r(x).
\end{equation}
Using Wiener-Khinchin theorem the correlation function $r(x)$ can be expressed in terms of the noise power spectrum $S(k)$
\begin{equation}\label{eq:n7}
    r(x) = \int \mathrm{d}k\,\, S(k) e^{ikx} = \int \mathrm{d}k\,\, S(k) \cos(kx).
\end{equation}
Substituting \ref{eq:n7} into \ref{eq:n6} and evaluating the integral over $x$ one obtains
\begin{equation}
    \sigma^2_L = \int \mathrm{d}k\,\, \sinc^2\left(\frac{kL}{2}\right) S(k).
\end{equation}
When applied to the real data this equation transforms into a sum over the finite number of modes, where $S_k$ can be calculated for example with the help of the Fast Fourier Transform (FFT). Thus, we arrive at the equation \ref{eq:sigma_L}:
\begin{equation}
    \sigma^2_\mathrm{L} = 2 \sum_{k=k_\mathrm{min}}^\infty \, \sinc^2\left(\frac{kL}{2}\right) S_k.
    \label{eq:sigma_L}
\end{equation}
In practice we calculate the power spectrum $S_k$ by making a FFT of the data strips of constant latitude binned into 360 bins by longitude $l$. The FFT is made after removal of outliers, as described in Appendix~\ref{appendix:cleaning}. An example of such transformation is shown in Fig.~\ref{fig:strip_fourier} for the RM data strip. In order to exclude the contribution of the regular field we take into account only the modes with $k\ge k_\mathrm{min}$, where $k_\mathrm{min}=3$ corresponds to the wavelength of $90^\circ$.

\section{Details of the final fit and comparison with existing models}
\label{appendix:comparison}
In this section we present a comparison of our model with the \texttt{base} model of \citet{Unger:2023lob}. In order to reproduce the results of \citet{Unger:2023lob} we switched to the thermal electron model of \citet{2017ApJ...835...29Y}, recomputed cosmic ray electron distribution following the prescriptions given in \citet{Unger:2023lob} and included the effect of the striated field. 

The bin by bin performances of the models are shown in Figs.~\ref{fig:bins_rm}~\ref{fig:bins_stokes_q}~\ref{fig:bins_stokes_u}. Our model and the model of \citet{Unger:2023lob} were fitted assuming different masks on the data. Apart from small scale features, the main difference between the masks comes from the Galactic plane and the Fan region. The Galactic plane in the RM data is excluded in our analysis, while present in the analysis of \cite{Unger:2023lob}. To the contrary, in the polarization data the Galactic plane is masked by \cite{Unger:2023lob} while included in our fits. Additionally, \cite{Unger:2023lob} masked the Fan region on the Stokes $Q$ and $U$ maps. Another two bright radio features Loop I and Loop II were excluded in both analysis in Stokes $Q$ and $U$. 

Summarizing the above, our mask is larger for the RM dataset which can be seen by the absence of data points in the lower panel of Fig.~\ref{fig:bins_rm}. On the other hand, the mask on the polarization data is larger in \citet{Unger:2023lob}. The additional regions masked out by \citet{Unger:2023lob} are roughly shown with grey shading in the Figs.~\ref{fig:bins_stokes_q} and \ref{fig:bins_stokes_u}. Note that the model by \citet{Unger:2023lob} does not fit the data in these regions.

As one can see from Figs.~\ref{fig:bins_rm}~\ref{fig:bins_stokes_q}~\ref{fig:bins_stokes_u}, in general the predictions of the models are quite similar. However, the direct comparison of the models in terms of $\chi^2$ is not straightforward since they were fitter assuming different masks on the data. For cross-check purposes, the conservative approach would be to compare the models only in those regions where both were fitted. Reproducing roughly the mask of \citet{Unger:2023lob} and taking into account only those bins which were included in both analyses we find that the total chi-squared/ndf for our model is $\chi^2 = 1070/717$, while for the \texttt{base} model of \citet{Unger:2023lob} one finds $\chi^2 = 1436/717$. Most of the difference comes from the RM data.

\begin{figure}
    \centering
    \includegraphics[width=\linewidth]{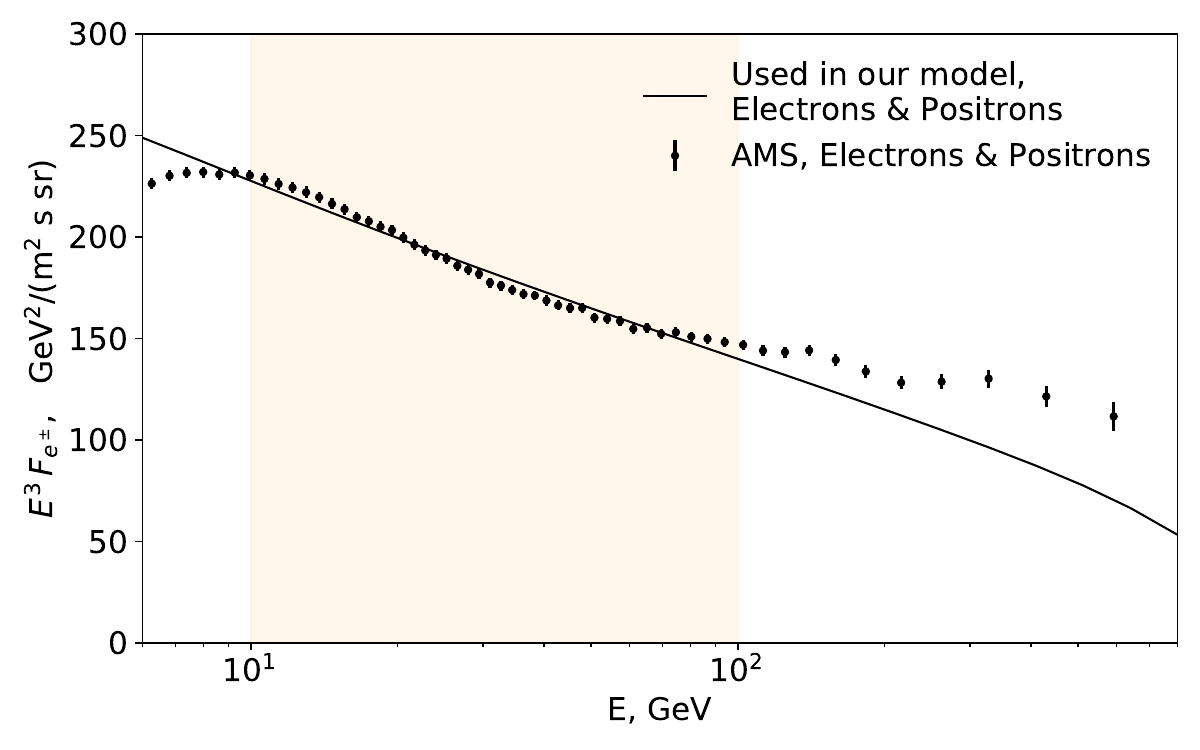}
    \caption{Spectrum of relativistic leptons. Black dots show the measurements by the AMS-02 experiment. The black curve corresponds to the CRE spectrum calculated with the DRAGON code as described in Sec.~\ref{sec:cr_elec} and used in the analysis. The CRE within the energies indicated with the yellow band produce more than 95\% of the synchrotron emission at 23~GHz. The deviation of the AMS-02 measurements below 10~GeV is due to the Solar modulation which we do not take into account.}
    \label{fig:CRE_AMS}
\end{figure}

\begin{figure*}
    \centering
    \includegraphics[width=0.8\linewidth]{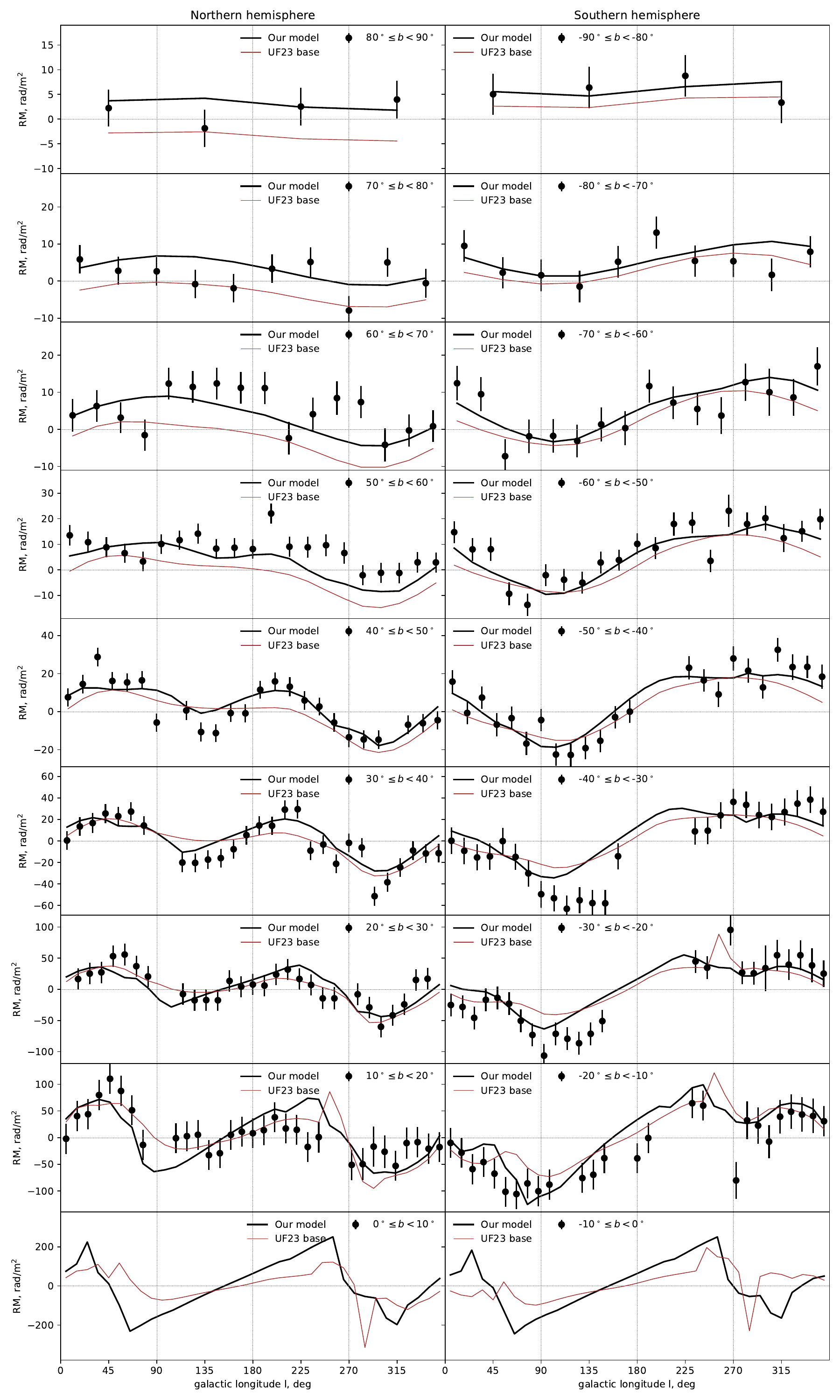}
    \caption{Our best fit model (black line) and UF23 \texttt{base} model (red line) in comparison with the binned RM data (black points with errorbars). }
    \label{fig:bins_rm}
\end{figure*}
\begin{figure*}
    \centering
    \includegraphics[width=0.787\linewidth]{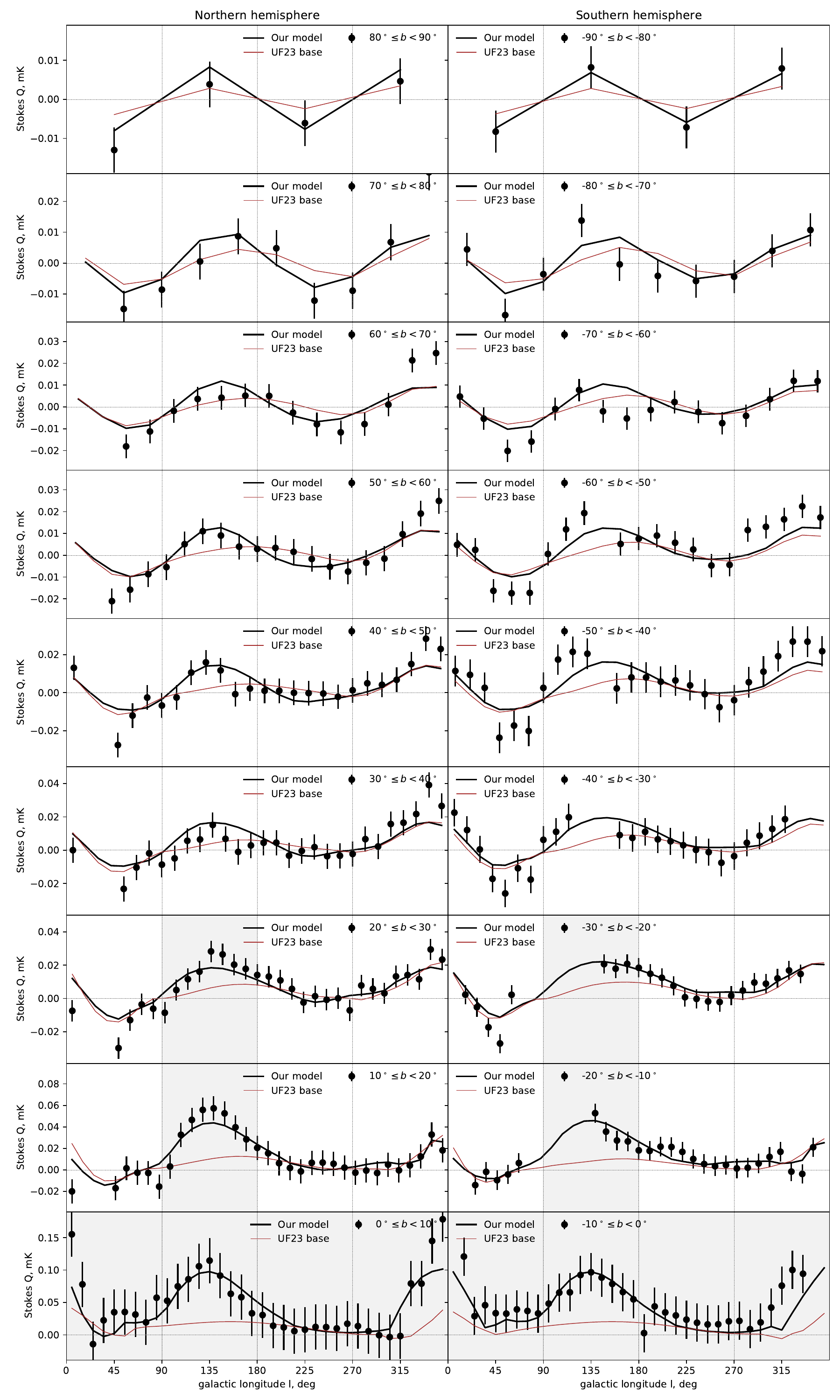}
    \caption{Same as Fig.\ref{fig:bins_rm} but for the Stokes Q parameter. The gray shading roughly indicates the regions which were excluded in the analysis of \citet{Unger:2023lob}.}
    \label{fig:bins_stokes_q}
\end{figure*}
\begin{figure*}
    \centering
    \includegraphics[width=0.8\linewidth]{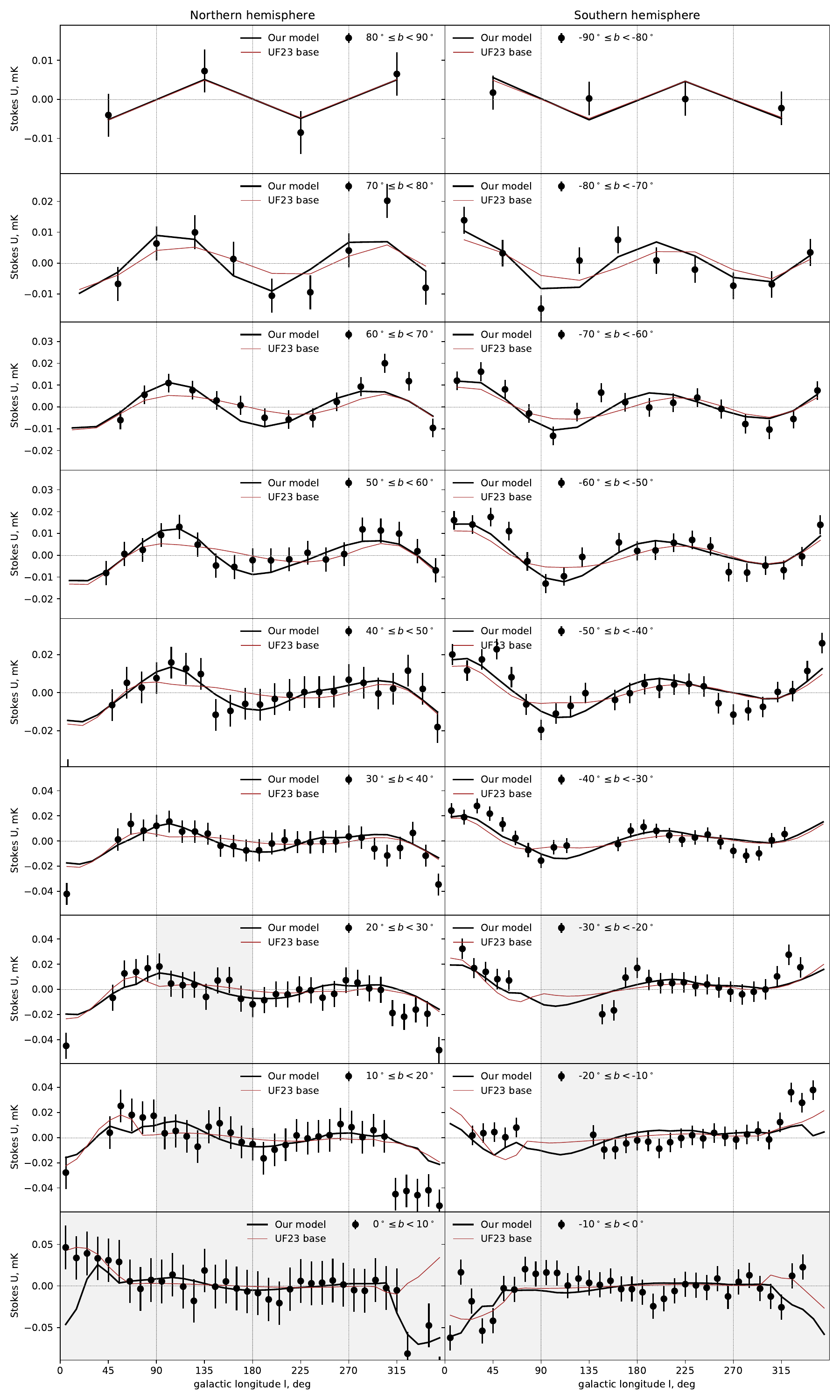}
    \caption{Same as Fig.\ref{fig:bins_rm} but for the Stokes U parameter.}
    \label{fig:bins_stokes_u}
\end{figure*}
\begin{figure*}
    \centering
    \includegraphics[width=0.97\linewidth]{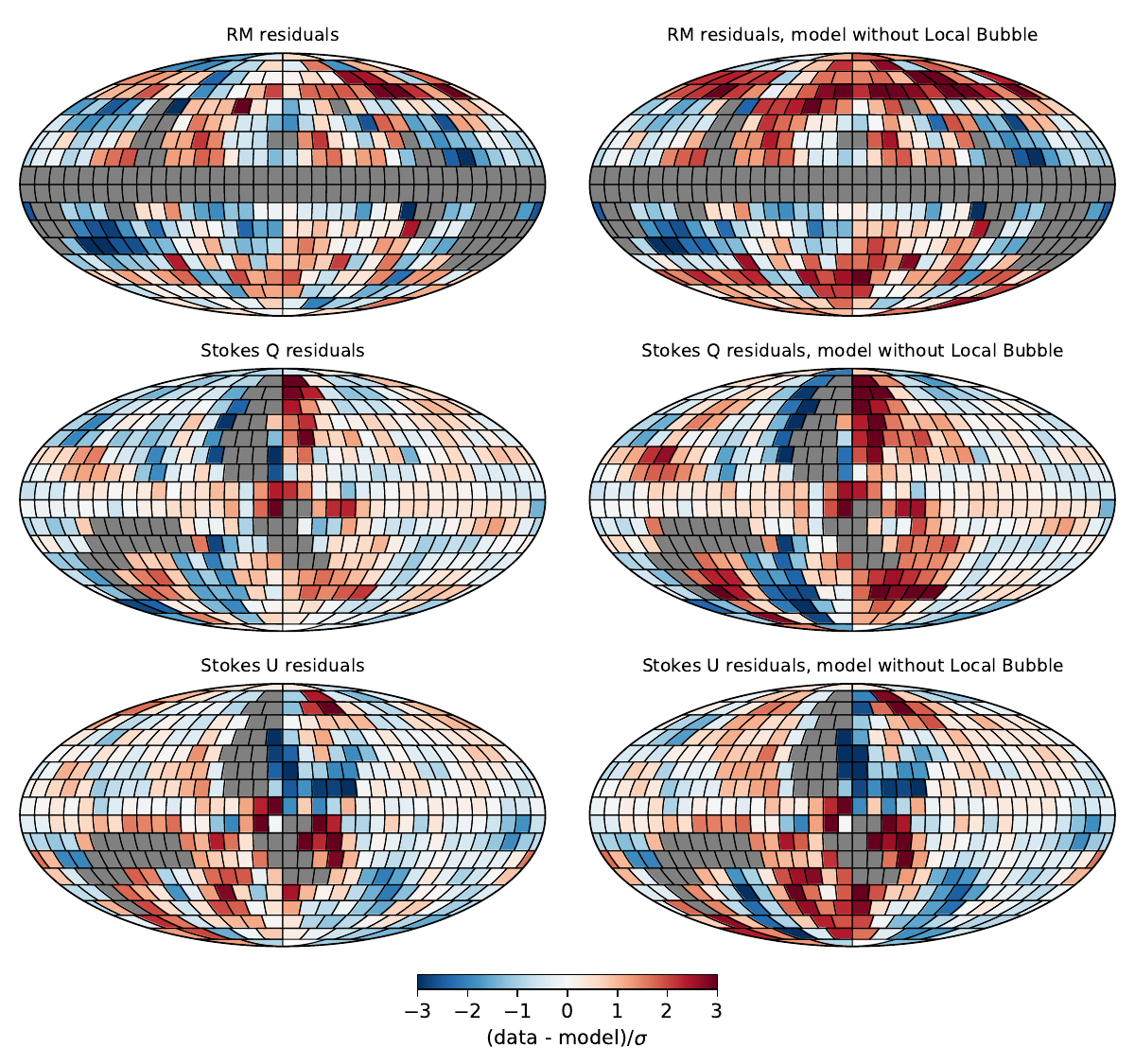}
    \caption{Difference between the data and the model in the units of bins errorbars. The colormap saturates when the difference reaches 3$\sigma$ level. Left column corresponds to the full model, while the right column shows the residuals for the model with the signal from the Local Bubble removed. The improvements due to the contribution of the Local Bubble are clearly visible in all observables, especially at high Galactic latitudes $|b| > 60^\circ$.}
    \label{fig:residuals}
\end{figure*}
\begin{figure*}
    \centering
    \includegraphics[width=0.97\linewidth]{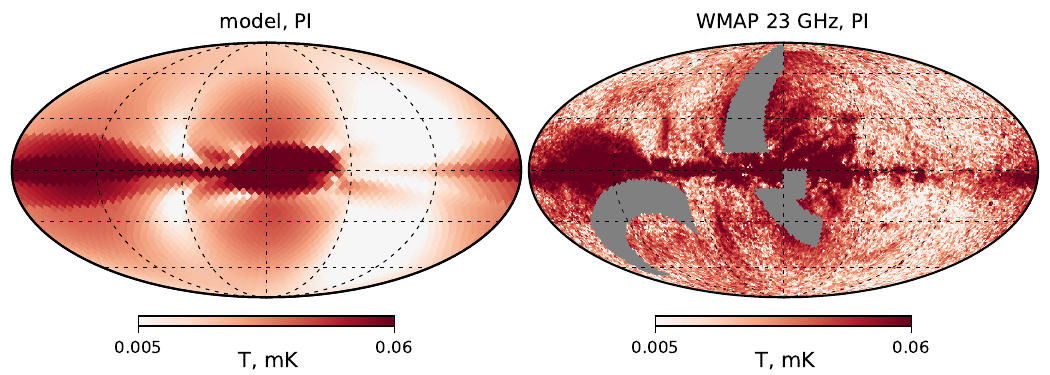}
    \caption{Polarized synchrotron intensity of the model in comparison with the WMAP 23~GHz data. The colomap is in the logarithmic scale.}
    \label{fig:PI_skymaps}
\end{figure*}

\end{appendix}
\end{document}